\documentclass[sigconf,authorversion,nonacm]{acmart}

\usepackage{multirow,makecell}
\usepackage{xcolor,colortbl}
\usepackage{pifont}
\usepackage{enumitem}
\usepackage{subfig}
\usepackage{graphicx}
\usepackage[flushleft]{threeparttable}

\usepackage{array} 
\usepackage[many]{tcolorbox}

\tcbset{
    frame code={},
    center title,
    left=0pt,
    right=0pt,
    top=0pt,
    bottom=0pt,
    colback=gray!50,
    colframe=white,
    width=\dimexpr\columnwidth\relax,
    enlarge left by=0mm,
    boxsep=0pt,
    arc=0pt,
    outer arc=0pt,
}

\usepackage{graphicx}
\usepackage{xspace}
\usepackage{multicol}

\usepackage{ifthen}
\usepackage[normalem]{ulem} 
\newboolean{showcomments}
\setboolean{showcomments}{true} 
\ifthenelse{\boolean{showcomments}}
  {

		\newcommand{\nbb}[2]{
		\fcolorbox{black}{red}{\textcolor{black}{\bfseries\tiny#1}}
		{$\blacktriangleright$\textcolor{red}{\textit{#2}}$\blacktriangleleft$}
		} 
		\newcommand{\nbc}[2]{
		\fcolorbox{black}{orange}{\bfseries\tiny#1}
		{$\blacktriangleright$\textcolor{blue}{\textit{#2}}$\blacktriangleleft$}
		}

		\newcommand{\nbf}[2]{
		\fcolorbox{black}{cyan}{\bfseries\tiny#1}
		{$\blacktriangleright$\textcolor{blue}{\textit{#2}}$\blacktriangleleft$}
		}

		\newcommand{\nbz}[2]{
		\fcolorbox{white}{red}{\textcolor{white}{\bfseries\tiny#1}}
		{\textcolor{red}{\textit{#2}}}
		}

	\newcommand{\remarks}[1]{\color{red}[#1]\color{black}}

	\newcommand{\del}[1]{\textcolor{red}{\sout{#1}}} 
		
	\newcommand{\removed}[1]{\cbstart\removedfragile{#1}\cbend{}}
	\newcommand{\removedfragile}[1]{{\color{red}{\sout{#1}}}{}}

	\newcommand\riccardo[1]{\nbf{riccardo}{#1}}
        
        \newcommand\hadi[1]{\nbc{hadi}{#1}}
        \newcommand\luca[1]{\nbc{luca}{#1}}
        \newcommand\vittoriano[1]{\nbf{vit}{#1}}

        \newcommand\todo[1]{\nbz{ToDo}{#1}}
        \newcommand\todoall[1]{\nbz{ToDo ALL}{#1}}

  }
  {
	\newcommand{\nbb}[2]{}
	\newcommand{\remarks}[1]{}

	\newcommand{\del}[1]{} 
		
	\newcommand{\removed}[1]{} 
  	\newcommand{\removedfragile}[1]{}

	\newcommand\riccardo[1]{}
        \newcommand\claudio[1]{}
        \newcommand\hadi[1]{}
        \newcommand\luca[1]{}
        \newcommand\vittoriano[1]{}

	\newcommand\todo[1]{}
	\newcommand\todoall[1]{}

	\newcommand\todohugo[1]{}
        \newcommand\todomassimo[1]{}
        \newcommand\todoabel[1]{}
        \newcommand\todojoan[1]{}
        \newcommand\todoantonio[1]{}
        \newcommand\todowasif[1]{}
        \newcommand\todofernando[1]{}
        \newcommand\todoluca[1]{}
        \newcommand\todogilles[1]{}
        \newcommand\todomanuel[1]{}
        \newcommand\todoromina[1]{}
        \newcommand\todovittoriano[1]{}
        \newcommand\todobilal[1]{}
        \newcommand\todoetienne[1]{}
        \newcommand\todolucap[1]{}
        \newcommand\todolaura[1]{}
        \newcommand\todomehrdad[1]{}
        \newcommand\todopasqualina[1]{}
        \newcommand\todosandra[1]{}
  }

\newcommand{\ext}[1]{}

\newcommand\aidoart[1]{AIDOaRt}

\begin{document}
%


\title{Towards Synthetic Trace Generation of Modeling Operations using In-Context Learning Approach}


\author{Vittoriano Muttillo}
\affiliation{%
  \institution{\textit{University of Teramo}}
  \city{Teramo}
  \country{Italy}}
\email{vmuttillo@unite.it}

\author{Claudio Di Sipio}
\affiliation{%
  \institution{\textit{University of L'Aquila}}
  \city{L'Aquila}
  \country{Italy}
}
\email{claudio.disipio@univaq.it}

\author{Riccardo Rubei}
\affiliation{%
  \institution{\textit{University of L'Aquila}}
  \city{L'Aquila}
  \country{Italy}
}
\email{riccardo.rubei@univaq.it}

\author{Luca Berardinelli}
\affiliation{%
  \institution{\textit{Johannes Kepler University}}
  \city{Linz}
  \country{Austria}
}
\email{luca.berardinelli@jku.at}

\author{MohammadHadi Dehghani}
\affiliation{%
  \institution{\textit{Johannes Kepler University}}
  \city{Linz}
  \country{Austria}
}
\email{mohammadhadi.dehghani@jku.at}
\renewcommand{\shortauthors}{Muttilo et al.}



%



\newcommand*{\ie}{i.e.,\@\xspace}
\newcommand*{\eg}{e.g.,\@\xspace}
\newcommand*{\cf}{cf.\@\xspace}
\newcommand*{\RM}{README\@\xspace}

\newcommand*{\MG}{MORGAN\@\xspace}
\newcommand*{\HP}{Hepsycode\@\xspace}

\newcommand*{\GH}{GitHub\@\xspace}


\newcommand{\code}[1]{{\footnotesize \texttt{#1}}}

\newcommand{\cmark}{\ding{51}}%
\newcommand{\xmark}{\ding{55}}%
\newcommand*\circled[1]{\tikz[baseline=(char.base)]{\color{black} 
		\node[shape=circle,draw=cyan,fill=black!10!white,inner sep=.3pt] (char) {\sffamily{\small{\textbf{#1}}}};}}
	
\def\checkmark{\tikz\fill[scale=0.4](0,.35) -- (.25,0) -- (1,.7) -- (.25,.15) -- cycle;} 

\makeatletter
\newcommand*{\etc}{%
	\@ifnextchar{.}%
	{etc}%
	{etc.\@\xspace}%
}
\makeatother
\newcommand*{\etal}{\emph{et~al.}\@\xspace}

\newcommand\revision[1]{\textcolor{blue}{#1}}

\newcommand{\rqfirst}{\textbf{RQ$_1$}: \textit{How similar are the traces generated by LLMs to those generated by humans?}}

\newcommand{\rqsecond}{\textbf{RQ$_2$}: \textit{To what extent LLM-based traces can be used to train an IMA?}}

\newcommand{\rqthird}{\textbf{RQ$_3$}: \textit{Can the proposed approach be useful for modeling assistance in real-world use cases exploited in EU projects?}}

\newcommand{\abox}[1]{
	\begin{tcolorbox}[leftrule=1mm,toprule=0mm,bottomrule=0mm,left=1pt,right=2pt,top=2pt,bottom=2pt]
		\em #1
		 #1
	\end{tcolorbox}
}

\definecolor{verylightgray}{gray}{0.92}

\begin{abstract}

Producing accurate software models is crucial in model-driven software engineering (MDE). However, modeling complex systems is an error-prone task that requires deep application domain knowledge. In the past decade, several automated techniques have been proposed to support academic and industrial practitioners by providing relevant modeling operations. Nevertheless, those techniques require a huge amount of training data that cannot be available due to several factors, \eg privacy issues.
The advent of large language models (LLMs) can support the generation of synthetic data although state-of-the-art approaches are not yet supporting the generation of modeling operations. 
To fill the gap, we propose a conceptual framework that combines modeling event logs, intelligent modeling assistants, and the generation of modeling operations using LLMs. In particular, the architecture comprises modeling components that help the designer specify the system, record its operation within a graphical modeling environment, and automatically recommend relevant operations. In addition, we generate a completely new dataset of modeling events by telling on the most prominent LLMs currently available. As a proof of concept, we instantiate the proposed framework using a set of existing modeling tools employed in industrial use cases within different European projects. To assess the proposed methodology, we first evaluate the capability of the examined LLMs to generate realistic modeling operations by relying on well-founded distance metrics. Then, we evaluate the recommended operations by considering real-world industrial modeling artifacts.
Our findings demonstrate that LLMs can generate modeling events even though the overall accuracy is higher when considering human-based operations. In this respect, we see generative AI tools as an alternative when the modeling operations are not available to train traditional IMAs specifically conceived to support industrial practitioners. 

\end{abstract}
\maketitle



%

\section{Introduction}
\label{sec:Introduction}
Model-driven engineering (MDE) encompasses creating conceptual models that aim to represent complex software systems~\cite{ramos2011model,brambilla2017model}. However, such activity is error-prone since it relies on the modeler's expertise. In addition, modeling real-world industrial scenarios requires a deep knowledge of the application domain, thus requiring time to produce a suitable abstraction of the target system. In this respect, several automated approaches, \ie intelligent modeling assistants (IMAs)~\cite{mussbacher_opportunities_2020,MOSQUERA2024107492}, have been proposed to automate MDE-related activities. Although they represent a valuable solution to assist non-expert users, they need a large amount of training data to produce accurate recommendations~\cite{di2023morgan,mussbacher_opportunities_2020}. 



In this paper, we use the term \textit{modeling operations} to refer to activities performed by users on graphical model editors to create model(s) compliant with a given metamodel. Modeling operations generate \textit{events} captured by a notification mechanism and suitably saved as \textit{traces}. 
In this respect, logging modeling events in \textit{traces} can be beneficial~\cite{Dehghani2023}. Real traces are a collection of modeling events generated by users while editing models in suitably extended (graphical) editors, thus representing a source of realistic information for the IMAs. Nevertheless, collecting many real traces is challenging. To obtain \textit{enough} training data for IMAs, a reasonably large user base should be capable of using a modeling language and its supporting editor. The latter must be equipped with modeling event recording capabilities. 
The need for domain-specific languages~\cite{malavolta2013}, the heterogeneous technical landscape of language engineering platforms both in commercial solutions and academic modeling solutions~\cite{iung_systematic_2020}, as well as intellectual property and privacy concerns on collecting and sharing traces, hinders the use of IMAs in industrial contexts.

Advanced generative AI models like large language models (LLMs) can produce data closely resembling human style. However, it is crucial to check the produced data to mitigate the creation of fictional details or incorrect assertions, also known as hallucination phenomena~\cite{10.1145/3571730}.
While LLMs have been used to generate generic textual data in software engineering context~\cite{gerosa2024can}, no approaches are available to generate modeling operations to our best knowledge.

In this work, we present a conceptual MDE framework that aims to assess the capabilities of LLMs to generate modeling operations, collect them as traces, and feed IMAs. In particular, we adopt the in-context learning approach~\cite{10.1145/2043932.2043989} using few-shots prompting to generate modeling operations given a graphical model editor capable of collecting modeling event traces. To validate our conceptual framework, we rely on state-of-the-art modeling components developed during cooperation among research and industrial partners within different European projects
Concerning the generation of synthetic modeling operations, we experiment with the four most popular models, \ie GPT3.5 and 4 \cite{ChatGPT}, LLama3 70B\cite{LLama3}, and Gemini \cite{Gemini}. We then rely on the Eclipse Modeling Framework (EMF)~\cite{steinberg2008emf} and research solutions for modeling event recording~\cite{Dehghani2023} and \textit{modeling assistance}~\cite{di2023morgan} purposes.

In particular, we aim to answer the following research questions:

\noindent \ding{228} \rqfirst{}
To answer this question, we assess the quality of the generated traces by applying well-founded similarity distance metrics. In addition, we assess the degree of hallucination by defining a novel metric tailored to the context of MDE systems. 

\noindent \ding{228} \rqsecond{} After selecting the best LLM, we trained an existing IMA, \ie \MG,~\cite{di2023morgan} with human-generated traces, synthetic ones, and a mix of both. In particular, the goal is to investigate to what extent synthetic traces can be used to replace human-generated ones in real-world scenarios

\noindent \ding{228} \rqthird{} To answer this question, we elicit the best configuration to employ MORGAN as IMA on a real-world industrial validation set, aiming at understanding to what extent synthetic traces can be used in a real context.


Our findings show that GPT-4 produces the most accurate results by minimizing the hallucination even though the produced traces lower the \MG's overall accuracy. Nonetheless, mixing humans and generated traces can represent an adequate compromise when training models are not available at all, \eg when those artifacts are protected by intellectual property. In addition, LLMs, including open-source ones, can generate valuable events in a few minutes, representing a faster way to support the training phase of IMAs based on automated techniques. 

\textbf{Contributions.} The main contributions of the paper are the following: 
%
    \textit{(i)} a conceptual framework to integrate modeling operations, generated events, and traces and IMA tools to recommend relevant modeling operations in the context of MDE systems,
    \textit{(ii)} an evaluation of synthetic datasets constructed with prominent LLMs using well-founded distance metrics,
    \textit{(iii)} a real-world evaluation by considering software models employed in the context of EU projects,
    and \textit{(iv)} a replication package to foster further research in this domain~\cite{replicationPackageASE}.
    


\section{Background and Related Work}
\label{sec:background-related}

We overviewed the state-of-the-art to enlist the closest approaches to our work. To the best of our knowledge, there is no similar approach that employs LLM to generate synthetic traces in the context of an MDE environment. Nevertheless, we describe the state-of-the-art of the works that are related to our internal components.

\textbf{Modeling Trace Recording.}
In~\cite{Dehghani2023}, Dehghani et al. presented a tool that captures user interaction events through the EMF notification API~\cite{steinberg2008emf}, namely Modeling Event Recorder (MER).
Our approach leverages the MER component to collect traces.
In~\cite{herrmannsdoerfer2010towards}, Herrmannsdoerfer et al. discuss a generic operation recorder for model evolution based on an operation metamodel. As MER, it reuses EMF Notifications but neglects compatibility with standards like XES.
In~\cite{brosch2012introduction}, the authors exploited the concept of operation recording to perform model versioning. In particular, they relied on the tool Operation Recorder~\cite{herrmannsdoerfer2010towards} previously introduced.


\textbf{Synthetic Data Generation Leveraging LLMs.}
The recent advancement of LLMs has motivated the exploration of this technology to generate synthetic datasets.
In~\cite{Fan2023}, Fan et al. survey the use of LLMs (e.g., GPT, Llama, AlphaCode) in software engineering, discussing their potential, challenges (e.g., hallucinations) and the importance of hybrid approaches combining software engineering methods with LLMs. The survey covers mostly code generation, software testing, maintenance, evolution, and deployment tasks. Our work contributes to the hybridization effort, i.e., towards the effective integration of LLMs as part of an overall SE process. As such, we did not find evidence of using LLMs for trace generation to train IMAs. 
%

\textbf{Synthetic Data Generation in MDE context.}
Data scarcity is a well-known problem in the Model-Driven community. In the years, different approaches have been developed to ease this daunting issue~\cite{pietsch_generating_2011}. 
MDE researchers have tried to mitigate this challenge by presenting model generation techniques and tools. These studies mainly utilized three approaches: clustering, grammar graph, and random.
The clustering approach usually classifies variable values and relationships between components. Then, an instance model is produced from each category to represent that category~\cite{perez_test_2014}. To generate the model based on graph grammars, the graph rules are extracted from the metamodel, and then models are generated according to these rules~\cite{sen_mutation-based_2006}. In the random approach, random procedures are used to generate new models~\cite{mougenot_uniform_2009}. 
Machine learning techniques are also adopted for model generation.
In~\cite{lopez_generating_2022}, the authors have presented a deep learning-based framework for generating structurally realistic models.
In~\cite{rahimi_towards_2023}, Rahimi et al. used generative adversarial networks for generating new structurally realistic EMF models.
Notable tools for model generations in MDE relying on EMF~\cite{steinberg2008emf} are Wodel~\cite{gomez-abajo_wodel_2016} and Viatra~\cite{semerath_viatra_2019}. Wodel is conceived to generate seed model mutants. In particular, taking as input a seed model and a defined set of mutation rules, the tool can generate \textit{n} mutants. Viatra translates an EMF metamodel into a logic problem, which is solved by an underlying first-order logic solver to create valid graph models.
%
Cuadrado~\cite{cuadrado2020towards} conceived a tool to assist developers in creating test cases. In particular, the tool analyses the current version of the ATL transformation and derives a possible set of missing elements.

\textbf{Existing IMAs in MDE context.}
Several approaches have been developed to assist modelers in their daily tasks, \eg model transformations~\cite{cuadrado2018anatlyzer}, model repair~\cite{iovino2020model}, and model search~\cite{lopez2020mar}. Above all, model completion is the most supported task, leveraging on NLP techniques~\cite{burgueno2022generic,Ibrahimi2022}, similarity-based algorithms~\cite{di2022memorec,adhikari_simima_2023} and pre-trained models~\cite{weyssow2022recommending,chaaben2023towards} to recommend missing modeling elements given an incomplete model. However, none of the mentioned tools supports the recommendations of modeling operation apart from NEMO~\cite{di2022finding}, an IMA that forecasts the next modeling operations leveraging the LSTM network by relying on a curated dataset of BPMN models. However, we cannot reuse NEMO in a direct comparison since it is tailored for BPMN models. In this paper, we opt for \MG~\cite{di2023morgan},  an IMA that relies on graph-kernel and NLP since \textit{(i)} it has already been used to assist the completion of industrial context~\cite{cederbladh2024towards} and \textit{(ii)} the replication package is publicly available~\cite{replicationPackageMORGAN}.








\section{Motivating example}
\label{sec:motivationalexample}

\begin{figure}
    \centering
    \includegraphics[width=1.0\linewidth]{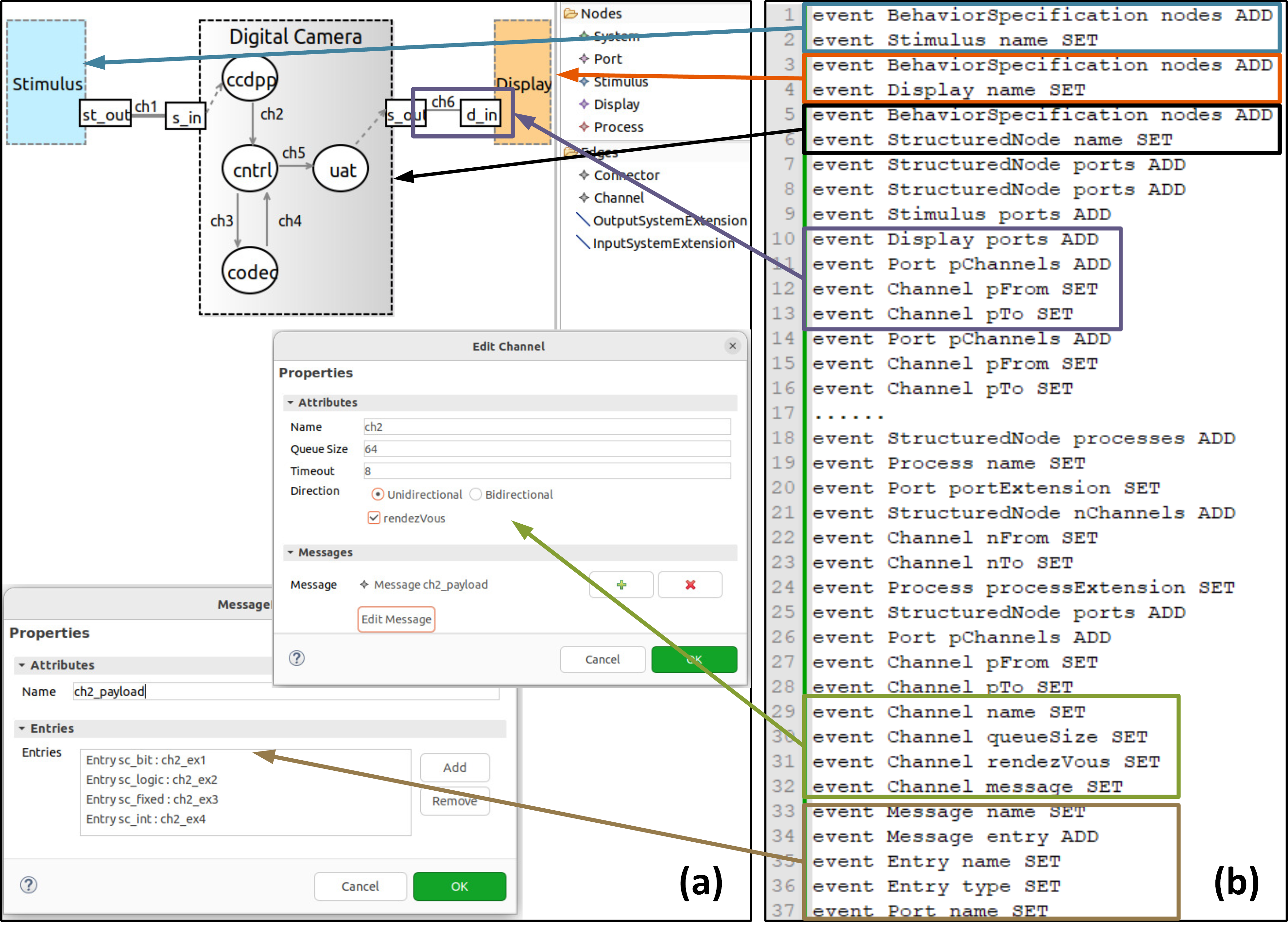}
    \caption{HEPSYCODE Graphical Modeling Workbench (a) and trace file generated through MER tool (b). The application considered in this scenario is called \textit{Digital Camera}~\cite{muttillo2023}.}
    \label{fig:hepsy_motivation}
\end{figure}

To model high-performance embedded systems, dedicated tools belonging to \textit{Electronic Design Automation} (EDA)~\cite{Huang2021}. With advancements in EDA, new methods and tools have emerged, enabling higher abstraction models through MDE approaches.
HEPSYCODE
(HW/SW \textbf{CO}-\textbf{DE}sign of \textbf{HE}terogeneous \textbf{P}arallel dedicated \textbf{SY}stems)~\cite{hepsycode-web} is a prototype EDA methodology and tool designed to reduce the design time of embedded applications. It uses Eclipse MDE technologies to model the behavior of embedded applications with a custom modeling workbench compliant with the HEPSYCODE metamodel. The HEPSYCODE language allows modeling the system as a network of processes communicating through channels.
Figure~\ref{fig:hepsy_motivation} (a) shows the HEPSYCODE graphical modeling workbench and the model of an embedded application called Digital Camera (DC). The main functionalities of the application, representing a camera that captures photographs in digital memory~\cite{vahid_1999}, include acquiring a 64×64-pixel image (i.e., \textit{ccdpp} process), performing a zero-bias adjustment (i.e., \textit{cntrl} process), compressing the image (i.e., \textit{codec} process), and transmitting it to an external device (i.e., \textit{uat} process). Data are exchanged through internal channels, while testbench and output feedback use additional external channels and ports. The metamodel defines various classes and entities used to model embedded applications as networks of processes, with data exchanged through channels and messages.

To assist designers, the modeling environment can collect their modeling operations as traces, as shown in Figure~\ref{fig:hepsy_motivation} (b). These traces can help IMAs suggest possible modeling operations at any given step, supporting designers as the complexity of the model increases.
Although IMAs can help modelers, several issues need to be carefully handled. Among the others, we elicited the following challenges:

\noindent \ding{228} \textbf{CH1: Collecting traces is time-consuming: } As shown in this section, collecting modeling operations is time-consuming if the system exploits a traditional event recorder. Moreover, the time needed to produce valuable traces depends strongly on the modeler's level of expertise, given the application domain. 

\noindent \ding{228} \textbf{CH2: Using external training data may lead to inaccurate results : } 
Alternatively, curated modeling datasets~\cite{modelset,babur2019metamodel} can be used to synthesize traces to feed IMA as done in~\cite{di2022finding}. Nonetheless, the existing datasets comprise models (and metamodels) specifically created for ML tasks without representing realistic modelers' behavior. 

\noindent \ding{228} \noindent \textbf{CH3: Accessing data is often difficult due to security issues and industry restrictions.} Industries and research institutions can handle data differently, according to internal or external regulations ~\cite{GAROUSI2016106}. This may impact modeling artifacts available to enable automated approaches, resulting in a scarcity of training data. In particular, privacy agreements can negatively affect data disclosing~\cite{10.1145/2647648.2647655}, thus preventing researchers from developing accurate automated approaches to support industrial practitioners. 











\section{Proposed approach}
\label{sec:Methodology}
\begin{figure}[ht!]
    \centering
    \includegraphics[width=1\linewidth]{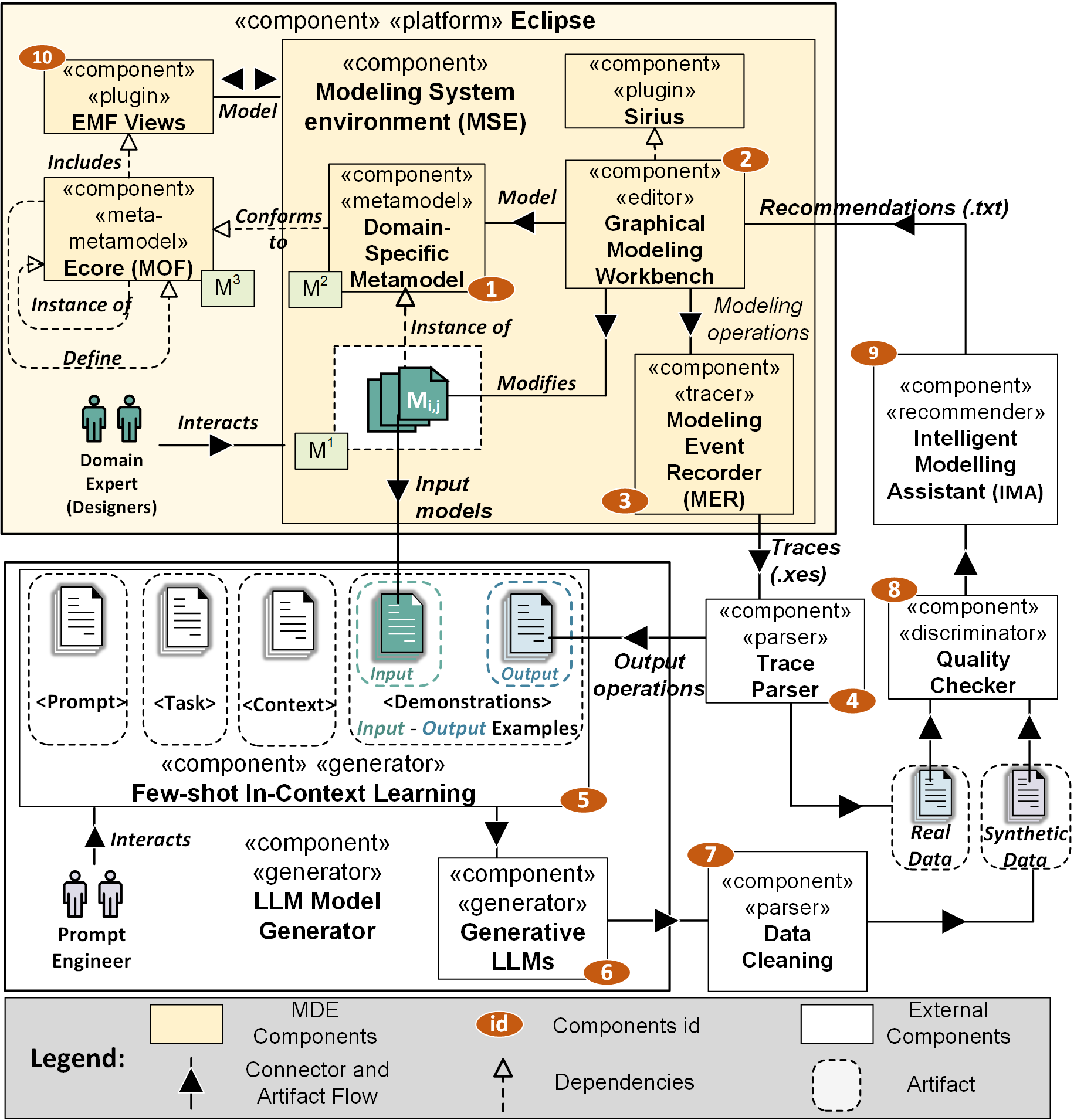}
    \caption{The proposed approach.}
    \label{fig:MER-MORGAN_approach} 
\end{figure}

This section introduces our approach. The core concept involves integrating modeling event recording capability within a modeling system environment. The goal is capturing events generated by users' modeling operations and generating traces. Such traces are then injected into recommender systems, enabling the generation of personalized suggestions for modeling actions most relevant to each designer. Components, connectors, and input/output artifacts flows are depicted in Figure~\ref{fig:MER-MORGAN_approach}. 

\subsection{Framework components}

This section details the framework's components as reported in Figure~\ref{fig:MER-MORGAN_approach}, referenced in the text using an \circled{id}, and their capabilities. It explains how they are integrated and used to support domain experts by suggesting modeling operations using IMAs trained with traces as LLM synthetic data. Section~\ref{sec:Evaluation} introduces a concrete implementation of such components.

\noindent\textbf{Modeling System Environment (MSE).}
The MSE is a graphical model editor. MDE leverages models as core software artifacts. The Meta Object Facility (MOF)~\cite{omgMetaObjectFacility}  by the Object Management Group (OMG) organizes modeling artifacts into three metalayers. Models ($M^1$) conform to a metamodel ($M^2$), which defines the modeling concepts and relationships of domain-specific languages~\circled{1}. Analogously, each metamodel conforms to meta-metamodel ($M^3$), which defines the concepts to define metamodels. 
MOF defines four model levels, with each level describing the one below. Building upon MOF, a modeling framework provides the necessary characteristics for working with these modeling artifacts
%
EMF~\cite{gronback_eclipse_2021} is a standard de-facto implementation of the MOF architecture. Its metamodeling language, Ecore, represents the $M^3$ layer. Using Ecore, developers define custom metamodels ($M^2$).

Formally, we define the set of L metamodels compliant with the Ecore as follows:
\begin{equation}\label{eq:eq1EE}
\begin{aligned}
    M^2 &= \{M^2_1, M^2_2, \cdots, M^2_i, \cdots,  M^2_L \}
\end{aligned}
\end{equation}
These metamodels then facilitate the automatic generation of graphical editors~\circled{2} for creating $M^1$ artifacts as instances of the conforming metamodel $M^2$. Formally, we also define the set of \textit{M} models conforming to Ecore-based metamodel $M^2_i$ as follows:
\begin{equation}\label{eq:eq1FF}
\begin{aligned}
    M^1_i &= \{M^1_{i,1}, M^1_{i,2}, \cdots, M^1_{i,j}, \cdots,  M^1_{i,M} \}
\end{aligned}
\end{equation}
%

\noindent\textbf{Modeling Event Recorder (MER).}
Formally, \textit{modeling operations} are activities performed by users on the Graphical Modeling Workbench to create model $M^1_{i,j}$ compliant with metamodel $M^2_i$. Modeling operations generate \textit{events} captured by a notification mechanism and suitably saved as \textit{traces} by a Modeling Event Recorder~\circled{3}.
Moreover, $\Gamma(M^1_i)$ is the set of traces obtained from modeling operations that realized the models $M^1_{i}$ conforming to $M^2_i$, such as:
\begin{equation}\label{eq:eq1GG}
\begin{aligned}
    \resizebox{0.9\hsize}{!}{$ \Gamma(M^1_i) = \{ \tau_1(M^1_{i,1}), \tau_2(M^1_{i,2}), \cdots, \tau_j(M^1_{i,j}), \cdots, \tau_m(M^1_{1,M}) \} $}
\end{aligned}
\end{equation}

To simplify matters, we will remove the internal model notation $\tau_j(M^1_{i,j})$ and only keep $\tau_{j}$ as a generic trace and $\Gamma_i$ as the set of traces. Each trace $\tau_{j}$ can be split into N events (i.e., single designer modeling operation), as follows:
\begin{equation}\label{eq:eq1HH}
\begin{aligned}
    \tau_{j} := \{ e_{j,1}, e_{j,2}, \cdots,  e_{j,k}, \cdots, e_{j,N} \}
\end{aligned}
\end{equation}

Each trace event has a fixed syntax, determined by the MER component:
\begin{equation}\label{eq:eq1II}
\begin{aligned}
    \resizebox{0.9\hsize}{!}{$ e_{j,k} := \text{event} \ <\text{class}> \ <\text{featureName}> \ <\text{eventType}> $}
\end{aligned}
\normalsize
\end{equation}


With the term modeling event recording, we refer to the collection of modeling traces through modeling event notification mechanisms. 


In the scope of the paper, the traces contain the timestamp of each operation. Thus, we consider this as a possible temporal relationship. While we consider including this kind of temporal relationship in our work, we notice that the recorded modeling operations are sequential, i.e., the recommender system is aware of this kind of sequence. In future works, we plan to investigate this interesting topic in depth.

\noindent\textbf{Intelligent Modeling Assistant (IMA).}
As defined in~\cite{mussbacher_opportunities_2020}, the development of an IMA encompasses the definition of several components. First, a \textit{data acquisition layer} must be defined to collect the relevant knowledge from external sources. In addition, an IMA operates in a \textit{context} where modelers perform their activities, thus producing contextual information that can be processed by the IMA. The core component is represented by the \textit{assistant}~\circled{9}, namely the algorithm used to perform the actual automated activities, \eg suggesting missing elements, retrieving similar modeling artifacts, or forecasting the next operations. An optional \textit{adaptation} phase of the IMA can be devised considering the modeler's feedback once the recommendations have been delivered. In the scope of the paper, we focus on IMAs that can retrieve relevant modeling operations given a graphical modeling environment. Formally, given the modeler's context (i.e., a model $M^1_{i,j}$), the knowledge acquired from external sources or the modeling context (i.e., $\Gamma(M^1_i)$ traces set), and \( A \) the assistant, the IMA is a function \( \mathcal{I} \) defined as follows:
\begin{equation}\label{eq:eq1AA}
\begin{aligned}
     \mathcal{I} (M^1_i,\Gamma(M^1_i),A) &=\{ Op_{i,1}, Op_{i,2}, \cdots, Op_{i,N} \}
\end{aligned}
\end{equation}
%


In the scope of the paper, we consider the \textit{past operations} $\Gamma_i$ as the unique source of knowledge for the IMA. Given the abovementioned definition of an event, an explanatory list of recommendations $Rec$ for a given model $(M^1_{i,j})$ is represented below:

\begin{equation}\label{eq:eq1BB}
\begin{aligned}
    Rec(M^1_{i,j}) = \{R(e_{j,1}), R(e_{j,2}), \cdots,  R(e_{j,N})\} 
\end{aligned}
\end{equation}
where the list  $\{R(e_{j,1}), R(e_{j,2}), \cdots,  R(e_{j,N})\} $ is the recommended modeling operations.





%
\begin{figure}
    \centering
    \includegraphics[width=1.0\linewidth]{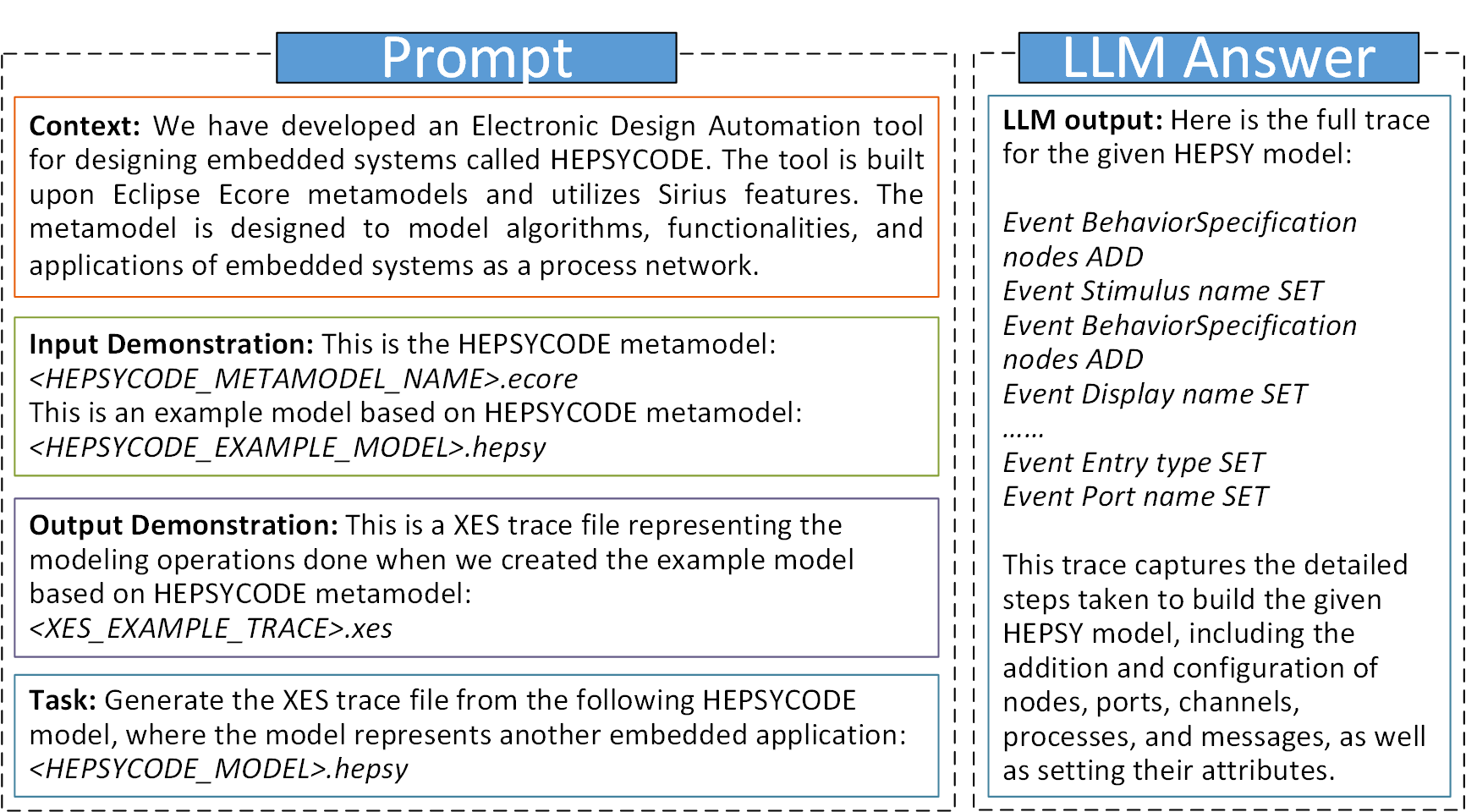}
    \caption{Prompt schema and LLM answer example.}
    \label{fig:MER-MORGAN_few_shot} 
\end{figure}

\textbf{Large Language Model (LLM).} A key phase of using LLMs is the design and execution of the most suitable prompt strategy given the goal~\cite{10.5555/3045118.3045347}. According to the recent literature, there are three main strategies:

\noindent\textit{Zero-shots} represents the basic prompt strategy, in which the LLMs are fed with just the query without any example of the expected outputs~\cite{10.5555/3045118.3045347}. The query can be expressed using natural language or embodying specific keywords that refer to the context,  e.g., parts of models under development.  

\noindent\textit{Few-shots} requires explanatory outputs in the initial query~\cite{LI2024112002}. In such a way, the task can be executed with weak labeling and minimal supervision from the developers. 

\noindent\textit{Chain-of-Thougts} is a conversational reasoning task to assess the ability of a model to maintain coherence and context across a series of questions and answers~\cite{10.1007/978-3-031-47994-6_24,10.1007/978-3-031-40292-0_1,DBLP:conf/nips/Wei0SBIXCLZ22}. Unlike traditional question answering tasks where each question is independent. 

In this paper, we adopt the \noindent\textit{few-shot} prompting strategy~\circled{5} to generate modeling operations~\circled{6} since it is suitable for obtaining the traces using the specified models. 
Furthermore, we defined $\Gamma^+(M^1_i)$ as the set of LLM synthetic traces (i.e., emulated human modeling operations) needed to realize the model $M^1_{i,j}$ conforms to $M^2_i$, such as:
\begin{equation}\label{eq:eq1CC}
\begin{aligned}
\resizebox{0.9\hsize}{!}{$
    \Gamma^+(M^1_i) = \{ \tau^+_1(M^1_{i,1}), \tau^+_2(M^1_{i,2}), \cdots, \tau^+_j(M^1_{i,j}), \cdots, \tau^+_M(M^1_{i,M'}) \}
$}
\end{aligned}
\end{equation}
Following modeling event recorder notation, we removed the internal model notation $\tau^+_j(M^1_{i,j})$ and only kept $\tau^+_{j}$ as a generic synthetic trace and $\Gamma^+_i$ as the set of synthetic traces. Each $\tau^+_{j}$ synthetic trace can be split into N synthetic events (i.e., single LLM human emulated modeling operation), as follows:
\begin{equation}\label{eq:eq1DD}
\begin{aligned}
    \tau^{+}_{j} &= \{ e^{+}_{j,1}, e^{+}_{j,2}, \cdots,  e^{+}_{j,k}, \cdots, e^{+}_{j,N'} \}
\end{aligned}
\end{equation}

The prompting engineering problem regards the creation of a $<Y> = \Gamma^{+}_{j}(M^1_{i})$ synthetic trace model dataset from a given set of models $M^1_i$ compliant with metamodel $M^2_i$, where LLM is fixed (no possible parameter tuning). The designer produces a generic model $<X> = M^1_{i,j}$ compliant with the considered metamodel $M^2_i$. We want to produce $<Y> = \tau^{+}(M^1_{i,j})$, i.e., a synthetic trace generated by LLM that tries to emulate actions needed by a human designer to generate model $M^1_{i,j}$ from specifications. The LLM is trained on demonstrations (i.e., input-output example traces) to emulate the human modeling operation steps using the Few-Shot In-Context Learning prompting approach~\circled{5}. Figure~\ref{fig:MER-MORGAN_few_shot} shows an example prompt schema used in this work.







\section{Evaluation Materials and Methods}
\label{sec:Evaluation}

This section details the evaluation materials and methods used in our work. Our approach ensures systematic analysis for reliable and reproducible results. We outline the employed tools, datasets, and evaluation approaches and explain the rationale behind their selection. 
All the tools and evaluation approaches were executed on a PC equipped with an Intel® Xeon CPU E3-1225 v5 @ 3.30 GHz, 32 GB system memory, 128KB LI cache, 1 MB L2 cache, and 8MB L3 cache. 

\subsection{Employed tools} \label{sec:toolchain}

\textbf{MSE Eclipse-based Workbench.}
The MSE is \textit{HEPSYCODE}~\cite{hepsycode-web} (HW/SW \textit{CO}-\textit{DE}sign of \textit{HE}terogeneous \textit{P}arallel dedicated \textit{SY}stems). It has been developed using the Eclipse EMF as the reference language workbench~\cite{iung_systematic_2020} and Sirius to generate its graphical modeling environment as a plugin of the Eclipse platform.
EMF~\cite{gronback_eclipse_2021}~\circled{10} is a standard de-facto implementation of the MOF architecture. Its metamodeling language, Ecore, represents the $M^3$ layer. Using Ecore, developers define custom metamodels ($M^2$).
The main component "Eclipse (<<platform>>)" in Figure~\ref{fig:MER-MORGAN_approach} represents our instantiation of the MOF architecture in which EMF and the Sirius plugin are used to create the Graphical Modeling Workbench~\circled{2}. The latter handles models ($M^1$) compliant with custom Domain Specific Metamodel~\circled{1} ($M^2$). Examples of this Graphical Modeling workbench can be found in~\cite{pomante2020} and~\cite{cederbladh2024towards}, which further extend the framework with EMF-compliant technologies (e.g., EMF Views~\cite{bruneliere2015emf}).
%
%
%
%
\textit{HEPSYCODE} is a framework and tool to improve the design time of embedded and CPS applications. It is based on a System-Level methodology for HW/SW Co-Design of Heterogeneous Parallel Dedicated Systems~\cite{muttillo2023}. HEPSYCODE uses Eclipse MDE technologies, SystemC custom simulator implementation, and AI-augmented algorithms for partitioning activities, all integrated into an automatic framework that drives the designer from the first input specifications to the final solution. 

The whole framework drives the designer from an \textit{Electronic System-Level} (ESL) behavioral model, with related Non-Functional (NF) requirements, to the final HW/SW implementation, considering specific HW technologies, scheduling policies, and Inter-Process Communication (IPC) mechanisms. The metamodeling language introduced in HEPSYCODE, named \textit{HEPSY}, is based on the Communicating Sequential Processes (CSP) Model of Computation (MoC). It allows modeling the system's behavior as a network of processes communicating through unidirectional synchronous channels. 
%
%
Moreover, a Model-to-Model (M2M) transformation involving the Xtext~\cite{xtext} framework has been used to translate the HEPSY model into an executable CSP-SystemC model of the system behavior. Through the execution of different simulation activities, including a system-level Design Space Exploration (DSE) approach that allows the related co-design methodology to suggest an HW/SW partitioning of the application specification and a mapping of the partitioned entities onto an automatically defined heterogeneous multi-processor architecture, it is possible to proceed with system implementation. The tool is freely available on GitHub.

\textbf{Modeling Event Recorder for EMF-based editors.}
The MER implementation for EMF-based models and editors is presented in~\cite{Dehghani2023} as subcomponents of a Modeling Process Mining Tool~\cite{Dehghani2023}~\circled{3}.
MER is thus implemented as an Eclipse plugin that interacts with Sirius-based graphical editors for EMF-based models, as the CAEX modeling
workbench~\cite{caex-workbench} or HEPSYCODE tool~\cite{hepsycode-web}, and records users’ modeling traces. Traces are encoded in the IEEE Standard for eXtensible Event Stream (XES)~\cite{XES}, which provides an XML schema for log encoding. The MER tool also provides an Ecore-based XES metamodel for encoding traces as EMF-based models.

\begin{figure}[h!]
    \centering
    \includegraphics[width=0.7\linewidth]{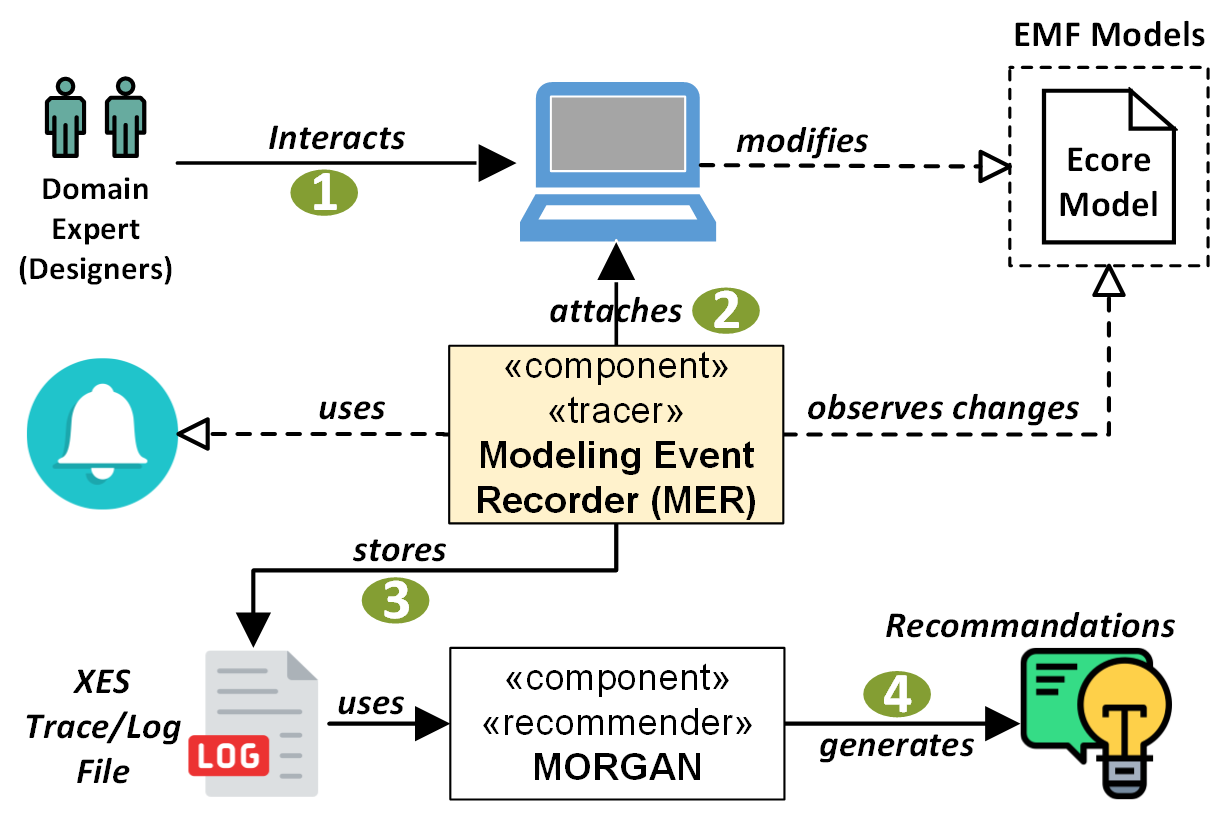}
    \caption{Modeling Event Recorder Workflow}
    \label{fig:mer_workflow} 
\end{figure}

Figure~\ref{fig:mer_workflow} shows the workflow in MER on a high level. The user starts interacting with the modeling editor. A modeling session starts when the editor is opened (step 1). Then, MER attaches itself to the created editing session (step 2) and listens to changes in the EMF-based models that are being modified by the graphical editors. As the user interacts with the editor and creates, modifies, and deletes graphical elements from the modeling canvas (e.g., internal elements, external interfaces, and internal links in the CAEX Modeling Workbench, or processes, channels, and ports in HEPSYCODE), the MER plugin captures all of the corresponding events. This change observation is done via the EMF Notification API, a built-in API in EMF. In step 3, as the editing session is closed (i.e., when the containing Eclipse project is closed),
the plugin stores the collected traces in an XES log file. Step 4 demonstrates that the MER output can be utilized as an input for a recommender, in this case, MORGAN. The recommender will then be able to generate recommendations based on these traces.

\textbf{MORGAN.}
As the IMA component, we consider the  \MG tool~\cite{di2023morgan} as discussed in Section~\ref{sec:background-related}.
Compared to the previous version~\cite{di2023morgan}, we modify \MG’s architecture by introducing a Trace parser~\circled{4} to extract relevant information from XES traces obtained by the MER component~\circled{3}~\cite{Dehghani2023}. In such a way, we obtain a textual-based representation used by the graph encoder to produce a list of trace graphs. To this end, the encoder extracts different features for each event, i.e., the type of event and the affected artifacts. Each graph is constructed using a set of Natural Language Preprocessing (NLP) techniques, i.e., stemming and dash-removal. Afterward, we assess the graph similarity between the XES graphs using the Weisfeiler-Lehman algorithm~\cite{shervashidze11a2011}, provided by the Grakel Python library~\cite{grakel-library}.
The outcomes of this Python component are the ranked list of similar operations given the context of the modeler, i.e., the initial XES trace.


\textbf{Large Language Model.}\label{LLM_subsec} LLM synthetic dataset generation uses pre-trained LLM to create synthetic datasets that mimic real-world data. This method leverages the extensive knowledge and contextual understanding embedded in LLMs to generate data that can be used for training, validation, and testing of various ML models. Furthermore, LLM synthetic dataset generation effectively addresses the challenges of reducing the time and effort to collect traces (CH1), representing realistic designer behavior (CH2), and overcoming security concerns, privacy regulations, and industry restrictions~\cite{10.1145/2647648.2647655}. By leveraging the advanced capabilities of LLMs, our proposed approach provides a robust solution for creating high-quality, relevant, and secure datasets, thereby enhancing the development and performance of ML models.

In such a context, our work used four LLMs~\circled{6} as instances of component \circled{6} in Figure~\ref{fig:MER-MORGAN_approach}:~(1) Gemini~\cite{Gemini}, developed by Google DeepMind;~(2) GPT-3.5~\cite{ChatGPT}, available for free and developed by OpenAI;~(3) GPT-4 Turbo \cite{ChatGPT4}, the professional version of GPT also developed by OpenAI;~(4) LLaMA3 70B~\cite{LLama3} developed by Meta. GPT-4, GPT-3.5, and Gemini are proprietary, meaning their underlying details and weights are not shared publicly. They are generally accessed via paid APIs or cloud services. LLama3 is an open-source model, offering transparency and flexibility with its publicly available architecture and weights. We selected these LLMs because they represent the best proprietary and open-source technologies according to LMSYS LeaderBoard~\cite{lmsys}.
%
%
In the scope of the paper, we used a web-based interface for all the selected LLMs. Therefore, we cannot configure the token size for each of them. For instance, GPT-4 offers a context window of 8K tokens. Noteworthy, we didn’t suffer from this limitation in the input prompts. Meanwhile, in the output phase, we experimented with this limitation in some cases. To cope with this, we asked the LLM to continue the generation phase in the same chat, thus preserving the active context.

\subsection{Datasets} \label{sec:datasets}

This section presents information about the dataset used for the training and evaluation steps of the proposed approach. In contrast, enhancement has been proposed with LLM in-context learning synthetic dataset generation.



\textbf{D1 - Hepsycode dataset:} The HEPSYCODE dataset is composed of embedded systems application models taken from the literature ~\cite{muttillo2023,rosvall2014,valente2021}, with a total amount of 2379 XES events,

%


\textbf{D2 - LLM datasets:} Starting from the models and trace files generated through the MSE flow, in this work we created a synthetic dataset using 4 LLMs available on the market, as presented in Section \ref{LLM_subsec}. For the creation of the datasets, we used the same prompt, as shown in Figure \ref{fig:MER-MORGAN_few_shot}, on all the considered four LLMs through the online query form.

\textbf{D3 - Industrial dataset:}
HEPSYCODE has also been used and validated through several industrial Use Case (UC) studies from European Projects, covering 4 domains (i.e., Avionics, Smart cities, Space, Automotive). The total amount of XES events for these industrial models is 1079, spanning over three different EU projects, \ie \textbf{MegaM@aRt$^2$}~\cite{megamart2}, \textbf{AQUAS}~\cite{aquas}, and \textbf{AIDOaRT}~\cite{aidoart-web}.


\subsection{Evaluating synthetic data} \label{sec:metrics}

Synthetic data generated from LLMs inherently faces several data limitations that must be acknowledged and addressed. As an inherent characteristic, LLMs may inadvertently propagate inaccuracies or biases present in their pre-training data, leading to outputs that may not always align with factual or unbiased information. Moreover, synthetic data generated by LLMs can sometimes not only be inaccurate but completely fictitious or disconnected from reality, a phenomenon often referred to as "hallucination".

To address these issues, the quality of the synthetic data can be assessed from the perspectives of diversity, correctness, and hallucination, measured using quantitative metrics, as presented below.

\subsubsection*{\textbf{Correctness Metric}} 

The \textit{Correctness} metric measures whether the data instance is related to the given label. Existing approaches for measuring correctness can be divided into two categories: automatic evaluation and human evaluation. Human evaluation has been conducted by prompt engineers to self-tune the \textit{Few-shot In-Context Learning} component. Automatic evaluation has been implemented to check the correctness of event syntax using the following metric:
\begin{equation}\label{eq:eq1}
\footnotesize
\begin{aligned}
    C(\tau^{+}_{j}) = \frac{\sum_{k=1}^{N'} c(e^{+}_{j,k})}{|\tau^{+}_{j}|}, \ \ \text{where} \ \
    c(e^{+}_{j,k}) = 
    \begin{cases}
          1 \ \text{if $e^{+}_{j,k}$ has correct syntax}  \\
          0 \ \text{otherwise}
    \end{cases}
\end{aligned}
\normalsize
\end{equation}
This metric can be evaluated on the full $\tau^{+}_{j}$ synthetic trace while it is possible to cluster the metrics w.r.t syntax features (i.e., MER metamodel classes).

\subsubsection*{\textbf{Diversity Metric}} \label{diversity_metrics}

\textit{Diversity} measures the difference between a chunk of text and another in the generated instances. In this work, we evaluate differences between $\tau^{+}_{j}$ synthetic traces generated by LLMs and real $\tau_{j}$ traces generated by designers using the MER component. The considered metrics are the follows: 

\textbf{Edit-based} similarities, also known as distance-based, measure the minimum number of single-character operations (e.g., insertions, deletions, or substitutions) required to transform one string into another. 

\textit{Levenshtein}: The Levenshtein distance $dist(\tau_{j},\tau^{+}_{j})$ between $\tau_{j}$ and $\tau^{+}_{j}$ is the minimum number of single-character edits (insertions, deletions, or substitutions) required to change one trace into the other. Starting from the Levenshtein distance, the Levenshtein similarity is defined as follows:
        \begin{equation}\label{eq:eq2}
        \resizebox{0.5\hsize}{!}{$
            \text{LEV}(\tau_{j},\tau^{+}_{j}) = 1.0 - \frac{dist(\tau_{j},\tau^{+}_{j})}{max(|\tau_{j}|,|\tau^{+}_{j}|)}
        $}
        \end{equation}
 \textit{Longest Common substrings (LCS)}: The maximum-length common events subsequence LCS(i,k) of $\tau_{j}$ and $\tau^{+}_{j}$, considering only characters insertion and deletion, where i and k represent the prefix length of trace string $\tau_{j}[i] \in \tau_{j}$ and $\tau^{+}_{j}[k] \in \tau^{+}_{j}$, respectively, is given by:
        \begin{equation}\label{eq:eq3}
        \resizebox{0.9\hsize}{!}{$
            LCS(i,k) = 
            \begin{cases}
                0 & \text{if} \ i = 0 \vee k = 0 \\
                LCS(i-1,k-1) + 1 & \text{if} \ i, k > 0 \wedge \tau_{j}[i] = \tau^{+}_{j}[k] \\
                0 & \text{if} \  i, k > 0 \wedge \tau_{j}[i] \neq \tau^{+}_{j}[k] 
            \end{cases} 
        $}
        \end{equation}
        \textit{Jaro–Winkler}: The Jaro Similarity is calculated using the following formula:
        \begin{equation}\label{eq:eq4}
        \resizebox{0.75\hsize}{!}{$
            \text{JARO}(\tau_{j},\tau^{+}_{j}) = 
            \begin{cases}
                0 & \text{if $m = 0$} \\
                \frac{1}{3} \left( \frac{m}{|\tau_{j}|} + \frac{m}{|\tau^{+}_{j}|} + \frac{m-t}{m} \right) & \text{Otherwise}
            \end{cases}
        $}
        \end{equation}
        where m is the number of matching characters between $\tau_{j}$ and $\tau^{+}_{j}$ and t is half the number of transpositions.
    
Among the \textbf{token-based} similarity function, we consider:

\textit{Jaccard}: measure the size of the intersection divided by the size of the union of the strings, as follows:
        \begin{equation}\label{eq:eq5}
            \text{JACCARD}(\tau_{j},\tau^{+}_{j}) = \frac{|\tau_{j} \cap \tau^{+}_{j}|}{|\tau_{j}| + |\tau^{+}_{j}| - |\tau_{j} \cap \tau^{+}_{j}|}
        \end{equation}
        \textit{Sorensen-Dice}: evaluate twice the number of elements common to both traces divided by the sum of the number of elements in each trace, as follows:
        \begin{equation}\label{eq:eq6}
            \text{DICE}(\tau_{j},\tau^{+}_{j}) = \frac{2 |\tau_{j} \cap \tau^{+}_{j}|}{|\tau_{j}| + |\tau^{+}_{j}|}
        \end{equation}
         \textit{Q-Gram}: count the number of occurrences of different q-grams in the two traces. Given a trace $\tau_{j}$ and let $v \in \Psi^q$ a q-gram, the total number of occurrences of v in $\tau_{j}$, denoted by G($\tau_{j}[v]$), is obtained by sliding a window of length q over the trace tokens. Given two traces $\tau_{j}$ and $\tau^{+}_{j}$, the Q-gram similarity is described as follows:
        \begin{equation}\label{eq:eq7}
            \text{Q-GRAM}(\tau_{j},\tau^{+}_{j}) = 1 - \frac{\sum\limits_{v \in  \Psi^q}|G(\tau_{j})[v] - G(\tau^{+}_{j})[v]|}{\sum\limits_{v \in  \Psi^q} max(G(\tau_{j})[v],G(\tau^{+}_{j})[v])}
        \end{equation}
        the traces are the closer relatives the more they have q-grams in common. 

\textit{Cosine}: similarity between two non-zero vectors of an inner product space that measures the cosine of the angle between them, as follows:
    \begin{equation}\label{eq:eq8}
        \text{COSINE}(\tau_{j},\tau^{+}_{j}) = cos(\theta) = \frac{\tau_{j} \cdot \tau^{+}_{j}}{||\tau_{j}|| \cdot ||\tau^{+}_{j}||}
    \end{equation}
    where $\tau_{j} \cdot \tau^{+}_{j}$ is the dot product between the vector $\tau_{j}$ and $\tau^{+}_{j}$, and $||\tau_{j}||$ represents the Euclidean norm of the vector $\tau_{j}$. The resulting measure of similarity spans from -1, signifying complete opposition, to 1, indicating absolute identity. A value of 0 signifies orthogonality or decorrelation, while values in between denote varying degrees of similarity or dissimilarity. For text matching, the attribute vectors $\tau_{j}$ and $\tau^{+}_{j}$ are usually the term frequency vectors of the documents.

These metrics can be used to evaluate how well LLMs can emulate both the designer's modeling approach and patterns, as well as human-based modeling approaches.


\subsubsection*{\textbf{Hallucination Metric}} \label{hall_metric}

In the scope of the paper, we define the hallucination as the number of additional operations, namely \textit{non-realistic events}, generated compared to the human ones by specifying the following metric:
%
%
\begin{equation}\label{eq:eq9}
\begin{aligned}
\resizebox{0.9\hsize}{!}{$
    H(\tau_{j},\tau^{+}_{j})_{\text{<event>}} = \frac{\text{Number of Synthetic Events} \ e^{+}_{j,k} \ \text{of type <event>}}{\text{Number of Real Events} \ e_{j,k} \ \text{of type <event>}}
$}
\end{aligned}
\end{equation}
This metric can be evaluated on the full $\tau^{+}_{j}$ synthetic trace file and $\tau_{j}$ real trace file and also for all the considered DSL metamodel classes.
If these metrics are greater than 1, then the LLM produces an incorrect synthetic trace file (i.e., hallucination results, the LLM adds more classes than those present in the real trace model).

\subsection{Evaluating modeling recommendations} \label{sec:metricsRec}



Concerning the produced recommendations, we set up an automatic evaluation of accuracy metrics aiming at mimicking the modeler's behavior. In the scope of this paper, we define the true positive (TP) as the correct recommended operation, false positive (FP) as the wrong operation, and false negatives (FN) as the operations that should be included in the recommendations but actually are not. Given these definitions, we define Precision (PR), Recall (REC), and F1-score (F1) as follows: $ PR = TP / (TP + FP) $;  $ REC = TP / (TP + FN)$; and $ F1 = 2* PR * REC /(PR + REC)$. 

	
	

Concerning the assessed modeling operations, we evaluate the approach using two different parameters, \ie \textbf{context ratio (CR)} and \textbf{cutoff (CO)}. The first paramenter represents the number of operations captured at a certain timestamp, \ie past operations. In the scope of the paper, we mimic three different stages of models, i.e., early stage, medium, and almost complete, considering three thresholds, \ie  0.2, 0.5, and 0.8 of the original testing model. Similarly, we variate the number of recommended operations by setting the CO parameters 3,5 and 10 operations as thresholds.

By relying on these two parameters, we derived nine different configurations represented in Table \ref{tab:configs}. 
In such a way, we can analyze how the overall accuracy varies according to contextual information. For instance, configuration $C_{1.1}$ represents the situation where the system recommends few operations in an early stage of development, \ie C0=3 and CR=0.2, respectively. 
\begin{table}[h!]
    \centering
    \small
    \caption{Configurations for the accuracy evaluation}
    \begin{tabular}{|l|c|c|c|}
        \hline
        & CR=0.2 & CR=0.5 & CR=0.8 \\ \hline
        CO=3  &  $C_{1.1}$  & $C_{1.2}$   &  $C_{1.3}$  \\ \hline
        CO=5  &  $C_{2.1}$  &  $C_{2.2}$  &  $C_{2.3}$ \\ \hline
        CO=10 &  $C_{3.1}$  & $C_{3.2}$   & $C_{3.3}$   \\ \hline
    \end{tabular}
    \label{tab:configs}
\end{table}
Each configuration has been evaluated using the 5-fold cross-validation since it is a well-founded strategy to automatically evaluate ML-based recommender systems \cite{10.1007/978-3-642-27387-2_25}. In particular, we split the operations into train, test, and ground truth (GT) data by resembling the \MG original setting presented in \cite{di2023morgan}. The train traces are used to feed the underpinning graph kernel engine and are compared with the testing ones. We obtained the ground truth data by relying on the CR parameter to vary the number of already performed operations. Concretely, augmenting CR reduces the number of operations to be predicted, \ie the GT operations. We eventually use the test and GT data to compute the accuracy using the metrics presented in Section \ref{sec:metrics}. To avoid any bias in the evaluation, we randomize the testing and GT operations, thus assuming that there is no temporal relationship between them. 

In addition, we analyze the time required to perform \textit{i)} the loading of training traces and encoding them in graph-based format and \textit{ii)} the recommendation for all the testing operations.

\section{Results}
\label{sec:Results}

\textbf{Addressing $RQ_1$.}
To answer this research question, we evaluate the quality of the generated synthetic data using the four LLMs listed in Section \ref{LLM_subsec} and the evaluation approach presented in Section \ref{sec:metrics}. The quality of the synthetic data has been assessed from the perspectives of diversity, correctness, and hallucination, measured using quantitative metrics through the \textit{python-text-distance} library~\cite{Pytextdist}.
All the statistical analysis have been performed using Jamovi~\cite{jamovi}
software version 2.3.28.0.

We also calculate the total evaluation time by summing the time needed to complete each activity, from creating a domain-specific metamodel to generating a final dataset for training IMAs. Moreover, using the PC configuration presented in Section \ref{sec:Evaluation}, the total time for modeling and trace collection in MSE was approximately 150 minutes, while the time for generating models using the four LLMs and performing validation was approximately 15 minutes.

\textbf{Correctness evaluation:}
To assess the correctness of the generated data, we apply two different evaluations: human and automatic. 
Human evaluation has been conducted by prompt engineers in component \circled{5} in Figure \ref{fig:MER-MORGAN_approach}. The designers start by selecting input and output demonstrations and analyzing the MSE context and the task to complete (i.e., the creation of synthetic traces). Prompt refinements are applied until good quality results are achieved (i.e., the generated traces have the needed syntax, contain a good number of events, and the data do not have inaccurate or completely fictitious events). After a few minutes, it was possible to define the basic structure of the prompt through human validation, as shown in Figure \ref{fig:MER-MORGAN_few_shot}, and we proceeded to the human data cleaning activities in component \circled{7} (e.g., remove unwanted text and/or characters, standardize the format, and format the files to be used appropriately in the subsequent steps) and the automatic validation of the LLM generation results throughout component \circled{8} (i.e., the Quality Checker).

The automatic validation was performed following Eq.\ref{eq:eq1} in component \circled{8} (i.e., the Quality Checker). The results for all the LLMs show syntactic correctness $C(\tau^{+}_{j})$ greater than 99\% with p-values << 0.001 for all LLMs using the \textit{One Sample T-Test} with \textit{Null Hypothesis} $H_0: \mu$ < 99\% (i.e., mean minor than 99\%). Therefore, we can reject the \textit{Null Hypothesis} $H_0$ and we can assess that all the LLMs can accurately emulate, from a syntactic point of view, the generation of reference traces under the considered configuration.

\begin{figure}
    \centering
    \includegraphics[width=1\linewidth]{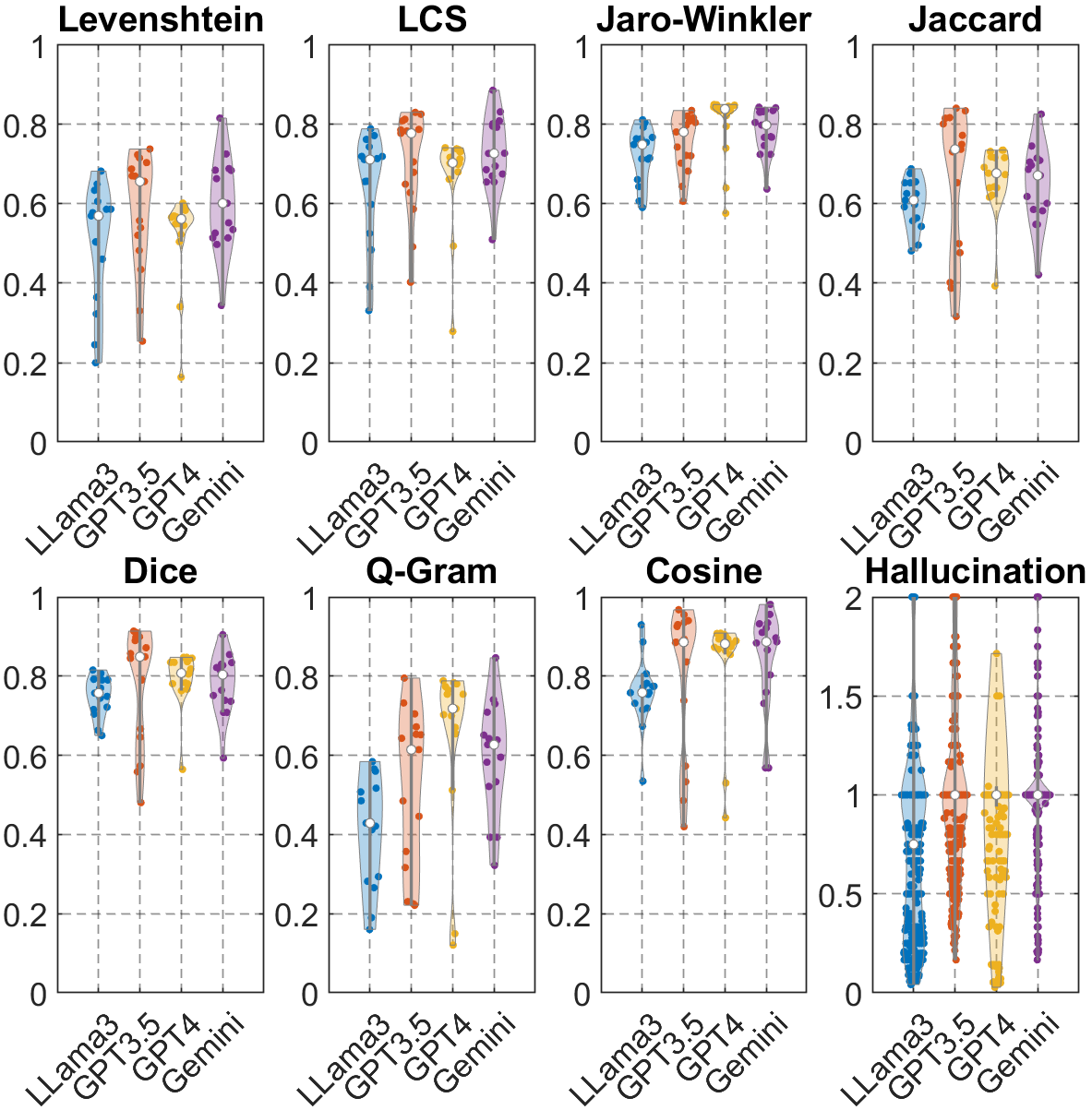}
    \caption{Synthetic Data Quality Evaluation Results. The violin plots show the distribution of points with the scatter plot. The white dots in the center represent the median.}
    \label{fig:metrics}
\end{figure}

\textbf{\textbf{Diversity evaluation:}} 
Concerning the diversity, we evaluate differences between $\tau^{+}_{j}$ synthetic traces generated by LLMs and real $\tau_{j}$ traces generated by designers using the diversity metrics defined in Section \ref{diversity_metrics}, from Eq. \ref{eq:eq2} to Eq. \ref{eq:eq8}. Figure \ref{fig:metrics} presents the violin plot results with the point distributions related to each considered metric value. The graph shows that all the LLMs behave similarly except for GPT-4, which has metric values closer to the median (i.e., lower variance). 

To identify significant differences among metrics, we conducted \textit{Welch's One-Way ANOVA} tests, which account for varying variances among LLM groups. This analysis reveals substantial differences in the JARO, COSINE, and Q-GRAM metrics across at least one LLM (p-values << 0.001). For a detailed analysis of metric group differences, we further utilized the \textit{Games-Howell} test for post-hoc analysis, where the pairs of LLMs with a p-value less than 0.05 are: \textit{(i)} \textit{JARO}: (GPT-4, GPT3.5) with p = 0.002; (GPT-4, Gemini) with p = 0.018; (GPT-4, Llama3) with p << 0.001. (ii) \textit{COSINE}: (GPT-4, Llama3) with p << 0.001. (iii) \textit{Q-GRAM}: (GPT-4, GPT3.5) with p = 0.004; (GPT-4, Gemini) with p = 0.010; (GPT-4, Llama3) with p << 0.001.


%
%
Finally, we used the \textit{Independent Samples T-Test} to determine which of the LLMs considered in the previous combinations is the best in terms of the mean value. This analysis confirms that GPT-4 outperforms the other LLMs in terms of the considered similarity measures.

\subsubsection*{\textbf{Hallucination:}} Automatic evaluation of non-realistic
events have been implemented considering the Hallucination metric defined in Eq. \ref{eq:eq9} in Section \ref{hall_metric}. Table \ref{tab:hallucination} shows the statistics related to the hallucination metrics. The Confidence Interval (CI) of the mean assumes sample means follow a t-distribution with N-1 degrees of freedom. 

\begin{table}[ht]
\centering
\caption{Hallucination metric result statistics.}
\resizebox{1.0\hsize}{!}{$%
    \begin{tabular}{lccccccccc}
        \toprule
        \textbf{LLM} & \textbf{N} & \textbf{$\mu$} & \textbf{SE} & \textbf{CI-L} & \textbf{CI-U} & \textbf{Median} & \textbf{SD} & \textbf{Variance} & \textbf{IQR} \\
        \midrule
        GEMINI & 330 & 0.969 & 0.0207 & 0.929 & 1.01 & 1 & 0.377 & 0.142 & 0.211 \\
        GPT35 & 307 & 1.066 & 0.0327 & 1.001 & 1.13 & 1 & 0.572 & 0.327 & 0.536 \\
        GPT4 & 335 & 0.912 & 0.0192 & 0.874 & 0.95 & 1 & 0.352 & 0.124 & 0 \\
        LLAMA3 & 306 & 0.775 & 0.0318 & 0.713 & 0.838 & 0.75 & 0.557 & 0.31 & 0.68 \\
        \bottomrule
    \end{tabular}
$%
} 
\begin{tablenotes}
\centering
\footnotesize
  \item $\mu$: Mean; SE: Standard Error; SD: Standard Deviation; IQR: Interquartile range; CI-L: 95\% Confidence Interval Lower Bound; CI-U: 95\% Confidence Interval Upper Bound; 
\end{tablenotes}
    \label{tab:hallucination}
\end{table}
%
%
Based on Table \ref{tab:hallucination}, GPT-4 is the only LLM with better values than the others, with a CI < 1 (CI-L = 0.874, CI-U = 0.95), an IQR equal to 0 (up to the 95th percentile all values are $\leq 1$, minimizing hallucination effects), and with SE (0.0192), SD (0.352) and Variance (0.124) lower than all other LLMs. Applying the \textit{One Sample T-Test} on GPT4 with the Null Hypothesis $H_0: (\mu_{GPT-4} > 1)$, we get a p-value << 0.001, allowing us to reject the $H_0$. This confirms that GPT-4, regarding the hallucination metric, is the best LLM among those considered.




\textbf{Answer to RQ$_1$:} The results show that the traces generated by the LLMs are very similar to those generated by a human, especially considering GPT-4, which has the best statistical results and reduced hallucination effects. This demonstrates that LLMs can be used as tools to reliably emulate modeling operations that a human could perform.

\subsubsection*{Addressing $RQ_2$} To answer this research question, we analyze two different aspects, \ie the amount of training data and how synthetic data can be used to replace human-generated operations. To support the first aspect, we run the 5-fold validation on the three datasets using the configurations discussed in Section \ref{sec:Evaluation}. Concerning the second aspect, we obtain a new dataset, \ie $D_2$, starting from the human-generated traces from dataset $D_1$ as discussed in Section \ref{sec:datasets}. To obtain $D_2$, we use the in-context few-shots learning setting described in Section \ref{sec:Methodology} by relying on GPT-4 LLMs, as it represents the best model according to the conducted evaluation in the previous research question. Moreover, we create the $D_{mix}$ dataset by randomly mixing traces belonging to $D_1$ and $D_2$.  
In addition, we analyze the recommendation capabilities in suggesting operations that affect \textit{i)} classes and \textit{ii)} attributes, resembling the original \MG experiment \cite{di2023morgan}. Table \ref{tab:class_results} shows the results obtained by \MG in recommending class operations considering the three datasets and the nine configurations.

It is evident that the system obtains better performance when real traces are used in all the considered configurations presented in Section \ref{sec:metrics}. In particular, configuration C3.3 leads to better performances, \ie the F1-score is equal to 0.60 on average. On the contrary, the maximum value achieved by \MG considering synthetic data is 0.34 using configuration $C_{3.2}$, meaning that the generated traces are less accurate compared to the human ones. 
Intuitively, powerful LLMs like GPT-4 can be used to generate modeling operations when real traces are not available. This claim is confirmed by analyzing the results obtained for $D_{mix}$ where the results are slightly increased compared to synthetic traces. On the one hand, we report that using only half of the real traces has limited impact. On the other hand, a small portion of synthetic data can contribute to enabling IMAs focused on recommending modeling operations. 
%
%
Concerning the configurations, we report that increasing both the CR and CO value contributes to increasing the overall performance of \MG, \ie the F1-score value is increased by 0.17 on average for all the datasets. 

\begin{table*}[h!]
    \centering
    \caption{Class recommendations results}
    \resizebox{\textwidth}{!}{
    \begin{tabular}{|c|c|c|c|c|c|c|c|c|c|c|c|c|c|c|c|c|c|c|c|c|c|c|c|c|c|c|c|c|c|c|c|c|c|c|c|c|c|c|c|c|c|c|c|c|c|c|c|c|c|c|c|c|c|c|c|}
        \hline
        & & \multicolumn{3}{c|}{$C_{1.1}$} & \multicolumn{3}{c|}{$C_{1.2}$} & \multicolumn{3}{c|}{$C_{1.3}$}
        & \multicolumn{3}{c|}{$C_{2.1}$} & \multicolumn{3}{c|}{$C_{2.2}$} & \multicolumn{3}{c|}{$C_{2.3}$}
        & \multicolumn{3}{c|}{$C_{3.1}$} & \multicolumn{3}{c|}{$C_{3.2}$} & \multicolumn{3}{c|}{$C_{3.3}$} \\ \hline
        & & PR & REC & F1 & PR & REC & F1 & PR & REC & F1 
        & PR & REC & F1 & PR & REC & F1 & PR & REC & F1 
        & PR & REC & F1 & PR & REC & F1 & PR & REC & F1 \\ \hline
        \multirow{6}{*}{\rotatebox{90}{Dataset $D_1$}} & R1 & 0.55 & 0.35 & 0.43 & 0.55 & 0.35 & 0.43 & 0.68 & 0.62 & 0.64 & 0.67 & 0.30 & 0.39 & 0.64 & 0.47 & 0.50 & 0.59 & 0.44 & 0.50 & 1.00 & 0.24 & 0.39 & 1.00 & 0.47 & 0.63 & 0.60 & 0.44 & 0.50 \\ \cline{2-29}
        & R2 & 0.62 & 0.42 & 0.49 & 0.62 & 0.42 & 0.49 & 0.58 & 0.59 & 0.57 & 0.63 & 0.31 & 0.34 & 0.65 & 0.40 & 0.45 & 0.68 & 0.61 & 0.63 & 1.00 & 0.24 & 0.38 & 0.96 & 0.40 & 0.56 & 0.73 & 0.57 & 0.63 \\ \cline{2-29}
        & R3 & 0.46 & 0.31 & 0.37 & 0.46 & 0.31 & 0.37 & 0.72 & 0.72 & 0.71 & 0.63 & 0.30 & 0.38 & 0.64 & 0.39 & 0.45 & 0.55 & 0.49 & 0.50 & 1.00 & 0.25 & 0.39 & 0.88 & 0.37 & 0.51 & 0.70 & 0.51 & 0.58 \\ \cline{2-29}
        & R4 & 0.49 & 0.39 & 0.42 & 0.49 & 0.39 & 0.42 & 0.59 & 0.68 & 0.60 & 0.63 & 0.28 & 0.37 & 0.76 & 0.49 & 0.57 & 0.69 & 0.63 & 0.64 & 1.00 & 0.22 & 0.36 & 0.84 & 0.39 & 0.53 & 0.76 & 0.64 & 0.68 \\ \cline{2-29}
        & R5 & 0.58 & 0.39 & 0.46 & 0.58 & 0.39 & 0.46 & 0.62 & 0.63 & 0.59 & 0.54 & 0.17 & 0.25 & 0.56 & 0.29 & 0.38 & 0.51 & 0.46 & 0.47 & 0.89 & 0.26 & 0.40 & 0.90 & 0.41 & 0.56 & 0.71 & 0.58 & 0.63 \\ \cline{2-29}
        & \cellcolor{verylightgray} Avg. & 
        \cellcolor{verylightgray}\textbf{0.54} 
        & \cellcolor{verylightgray}\textbf{0.37} 
        & \cellcolor{verylightgray} \textbf{0.43}
        & \cellcolor{verylightgray} \textbf{0.54}
        & \cellcolor{verylightgray} \textbf{0.37}
        & \cellcolor{verylightgray} \textbf{0.43}
        & \cellcolor{verylightgray} \textbf{0.64}
        & \cellcolor{verylightgray} \textbf{0.65} 
        & \cellcolor{verylightgray} \textbf{0.62} 
        & \cellcolor{verylightgray} \textbf{0.62} 
        & \cellcolor{verylightgray} \textbf{0.28} 
        & \cellcolor{verylightgray} \textbf{0.35}
        & \cellcolor{verylightgray} \textbf{0.65} 
        & \cellcolor{verylightgray} \textbf{0.41}
        & \cellcolor{verylightgray} \textbf{0.46}
        & \cellcolor{verylightgray} \textbf{0.61}
        & \cellcolor{verylightgray} \textbf{0.53}
        & \cellcolor{verylightgray} \textbf{0.55} 
        & \cellcolor{verylightgray} \textbf{0.98}
        & \cellcolor{verylightgray} \textbf{0.24}
        & \cellcolor{verylightgray} \textbf{0.38}
        & \cellcolor{verylightgray} \textbf{0.92}
        & \cellcolor{verylightgray} \textbf{0.40}
        & \cellcolor{verylightgray} \textbf{0.56} 
        & \cellcolor{verylightgray} \textbf{0.70}
        & \cellcolor{verylightgray} \textbf{0.54}
        & \cellcolor{verylightgray} \textbf{0.60} \\ \midrule

        \multirow{6}{*}{\rotatebox{90}{Dataset $D_2$}} & R1 & 0.44 & 0.27 & 0.30 & 0.32 & 0.27 & 0.29 & 0.23 & 0.34 & 0.27 & 0.33 & 0.10 & 0.15 & 0.31 & 0.17 & 0.21 & 0.32 & 0.33 & 0.30 & 0.78 & 0.15 & 0.25 & 0.63 & 0.24 & 0.34 & 0.36 & 0.35 & 0.33 \\ \cline{2-29}
        & R2 & 0.44 & 0.26 & 0.31 & 0.45 & 0.28 & 0.34 & 0.29 & 0.45 & 0.34 & 0.38 & 0.13 & 0.18 & 0.46 & 0.23 & 0.29 & 0.44 & 0.48 & 0.43 & 0.44 & 0.08 & 0.14 & 0.64 & 0.26 & 0.36 & 0.34 & 0.29 & 0.30 \\ \cline{2-29}
        & R3 & 0.36 & 0.21 & 0.25 & 0.25 & 0.17 & 0.20 & 0.28 & 0.43 & 0.32 & 0.33 & 0.16 & 0.19 & 0.38 & 0.15 & 0.22 & 0.31 & 0.26 & 0.28 & 0.78 & 0.15 & 0.25 & 0.56 & 0.21 & 0.31 & 0.38 & 0.36 & 0.35 \\ \cline{2-29}
        & R4 & 0.44 & 0.23 & 0.28 & 0.37 & 0.25 & 0.29 & 0.26 & 0.31 & 0.27 & 0.44 & 0.21 & 0.27 & 0.46 & 0.36 & 0.35 & 0.29 & 0.36 & 0.30 & 0.67 & 0.13 & 0.21 & 0.60 & 0.26 & 0.35 & 0.29 & 0.21 & 0.24 \\ \cline{2-29}
        & R5 & 0.33 & 0.15 & 0.20 & 0.43 & 0.28 & 0.33 & 0.31 & 0.42 & 0.34 & 0.48 & 0.21 & 0.26 & 0.44 & 0.24 & 0.30 & 0.28 & 0.31 & 0.28 & 0.78 & 0.15 & 0.25 & 0.55 & 0.21 & 0.30 & 0.55 & 0.40 & 0.44 \\ \cline{2-29}
& \cellcolor{verylightgray} Avg. & \cellcolor{verylightgray}\textbf{0.40} & \cellcolor{verylightgray}\textbf{0.24} & \cellcolor{verylightgray}\textbf{0.27} & \cellcolor{verylightgray}\textbf{0.38} & \cellcolor{verylightgray}\textbf{0.25} & \cellcolor{verylightgray}\textbf{0.29} & \cellcolor{verylightgray}\textbf{0.27} & \cellcolor{verylightgray}\textbf{0.39} & \cellcolor{verylightgray}\textbf{0.31} & \cellcolor{verylightgray}\textbf{0.39} & \cellcolor{verylightgray}\textbf{0.15} & \cellcolor{verylightgray}\textbf{0.21} & \cellcolor{verylightgray}\textbf{0.41} & \cellcolor{verylightgray}\textbf{0.23} & \cellcolor{verylightgray}\textbf{0.27} & \cellcolor{verylightgray}\textbf{0.33} & \cellcolor{verylightgray}\textbf{0.35} & \cellcolor{verylightgray}\textbf{0.32} & \cellcolor{verylightgray}\textbf{0.69} & \cellcolor{verylightgray}\textbf{0.13} & \cellcolor{verylightgray}\textbf{0.22} & \cellcolor{verylightgray}\textbf{0.60} & \cellcolor{verylightgray}\textbf{0.24} & \cellcolor{verylightgray}\textbf{0.34} & \cellcolor{verylightgray}\textbf{0.38} & \cellcolor{verylightgray}\textbf{0.32} & \cellcolor{verylightgray}\textbf{0.33}
\\ \midrule
        \multirow{6}{*}{\rotatebox{90}{Dataset $D_{m05}$}} & R1 & 0.33 & 0.11 & 0.14 & 0.27 & 0.20 & 0.22 & 0.38 & 0.47 & 0.41 & 0.48 & 0.26 & 0.30 & 0.40 & 0.30 & 0.31 & 0.34 & 0.30 & 0.31 & 0.72 & 0.19 & 0.29 & 0.51 & 0.27 & 0.34 & 0.46 & 0.40 & 0.41 \\ \cline{2-29}
        & R2 & 0.46 & 0.18 & 0.25 & 0.27 & 0.15 & 0.19 & 0.29 & 0.30 & 0.29 & 0.33 & 0.18 & 0.21 & 0.39 & 0.44 & 0.34 & 0.31 & 0.37 & 0.29 & 0.78 & 0.17 & 0.28 & 0.64 & 0.32 & 0.42 & 0.43 & 0.43 & 0.40 \\ \cline{2-29}
        & R3 & 0.48 & 0.13 & 0.20 & 0.43 & 0.37 & 0.36 & 0.34 & 0.43 & 0.36 & 0.39 & 0.15 & 0.21 & 0.38 & 0.27 & 0.29 & 0.35 & 0.50 & 0.38 & 0.72 & 0.15 & 0.25 & 0.58 & 0.26 & 0.35 & 0.39 & 0.37 & 0.36 \\ \cline{2-29}
        & R4 & 0.39 & 0.14 & 0.14 & 0.37 & 0.26 & 0.29 & 0.36 & 0.42 & 0.38 & 0.37 & 0.15 & 0.20 & 0.40 & 0.30 & 0.31 & 0.37 & 0.39 & 0.36 & 0.67 & 0.15 & 0.25 & 0.57 & 0.29 & 0.37 & 0.48 & 0.50 & 0.47 \\ \cline{2-29}
        & R5 & 0.46 & 0.26 & 0.28 & 0.33 & 0.25 & 0.28 & 0.36 & 0.41 & 0.38 & 0.43 & 0.17 & 0.22 & 0.44 & 0.23 & 0.29 & 0.43 & 0.40 & 0.41 & 0.78 & 0.20 & 0.31 & 0.60 & 0.29 & 0.38 & 0.34 & 0.32 & 0.32 \\ \cline{2-29}
              & \cellcolor{verylightgray} Avg. 
        & \cellcolor{verylightgray} \textbf{0.42} 
        & \cellcolor{verylightgray} \textbf{0.17} 
        & \cellcolor{verylightgray} \textbf{0.20} 
        & \cellcolor{verylightgray} \textbf{0.33} 
        & \cellcolor{verylightgray} \textbf{0.25} 
        & \cellcolor{verylightgray} \textbf{0.29} 
        & \cellcolor{verylightgray} \textbf{0.35} 
        & \cellcolor{verylightgray} \textbf{0.42} 
        & \cellcolor{verylightgray} \textbf{0.36} 
        & \cellcolor{verylightgray} \textbf{0.40} 
        & \cellcolor{verylightgray} \textbf{0.18} 
        & \cellcolor{verylightgray} \textbf{0.23} 
        & \cellcolor{verylightgray} \textbf{0.40} 
        & \cellcolor{verylightgray} \textbf{0.31} 
        & \cellcolor{verylightgray} \textbf{0.30} 
        & \cellcolor{verylightgray} \textbf{0.36} 
        & \cellcolor{verylightgray} \textbf{0.32} 
        & \cellcolor{verylightgray} \textbf{0.33} 
        & \cellcolor{verylightgray} \textbf{0.73} 
        & \cellcolor{verylightgray} \textbf{0.18} 
        & \cellcolor{verylightgray} \textbf{0.28} 
        & \cellcolor{verylightgray} \textbf{0.58} 
        & \cellcolor{verylightgray} \textbf{0.28} 
        & \cellcolor{verylightgray} \textbf{0.36} 
        & \cellcolor{verylightgray} \textbf{0.42} 
        & \cellcolor{verylightgray} \textbf{0.37} 
        & \cellcolor{verylightgray} \textbf{0.37} \\ \hline
    \end{tabular}}
    \label{tab:class_results}
\end{table*}






\begin{table*}[h!]
    \centering  
    \caption{Attributes recommendations results}
    \resizebox{\textwidth}{!}{
    \begin{tabular}{|c|c|c|c|c|c|c|c|c|c|c|c|c|c|c|c|c|c|c|c|c|c|c|c|c|c|c|c|c|c|c|c|c|c|c|c|c|c|c|c|c|c|c|c|c|c|c|c|c|c|c|c|c|c|}
        \hline
        & & \multicolumn{3}{c|}{$C_{1.1}$} & \multicolumn{3}{c|}{$C_{1.2}$} & \multicolumn{3}{c|}{$C_{1.3}$}
        & \multicolumn{3}{c|}{$C_{2.1}$} & \multicolumn{3}{c|}{$C_{2.2}$} & \multicolumn{3}{c|}{$C_{2.3}$}
        & \multicolumn{3}{c|}{$C_{3.1}$} & \multicolumn{3}{c|}{$C_{3.2}$} & \multicolumn{3}{c|}{$C_{3.3}$} \\ \hline
        & & PR & REC & F1 & PR & REC & F1 & PR & REC & F1 
        & PR & REC & F1 & PR & REC & F1 & PR & REC & F1 
        & PR & REC & F1 & PR & REC & F1 & PR & REC & F1 \\ \hline
        \multirow{6}{*}{\rotatebox{90}{Dataset $D_1$}} & R1 & 0.50 & 0.16 & 0.20 & 0.40 & 0.22 & 0.25 & 0.59 & 0.46 & 0.48 & 0.60 & 0.30 & 0.26 & 0.49 & 0.10 & 0.15 & 0.56 & 0.28 & 0.34 & 1.00 & 0.04 & 0.07 & 0.67 & 0.14 & 0.20 & 0.70 & 0.32 & 0.38 \\ \cline{2-29}
        & R2 & 0.52 & 0.20 & 0.26 & 0.46 & 0.21 & 0.26 & 0.60 & 0.45 & 0.46 & 0.47 & 0.10 & 0.14 & 0.55 & 0.25 & 0.26 & 0.52 & 0.37 & 0.36 & 1.00 & 0.04 & 0.08 & 0.88 & 0.16 & 0.24 & 0.71 & 0.23 & 0.34 \\ \cline{2-29}
        & R3 & 0.56 & 0.22 & 0.26 & 0.42 & 0.22 & 0.25 & 0.64 & 0.46 & 0.51 & 0.46 & 0.13 & 0.16 & 0.62 & 0.16 & 0.23 & 0.61 & 0.33 & 0.35 & 1.00 & 0.06 & 0.11 & 0.77 & 0.16 & 0.24 & 0.74 & 0.30 & 0.41 \\ \cline{2-29}
        & R4 & 0.61 & 0.27 & 0.32 & 0.48 & 0.30 & 0.32 & 0.57 & 0.36 & 0.42 & 0.71 & 0.18 & 0.21 & 0.69 & 0.20 & 0.26 & 0.74 & 0.35 & 0.43 & 1.00 & 0.03 & 0.05 & 0.90 & 0.07 & 0.13 & 0.88 & 0.33 & 0.41 \\ \cline{2-29}
        & R5 & 0.50 & 0.23 & 0.25 & 0.45 & 0.20 & 0.27 & 0.53 & 0.37 & 0.41 & 0.63 & 0.10 & 0.16 & 0.60 & 0.23 & 0.22 & 0.62 & 0.34 & 0.36 & 1.00 & 0.05 & 0.09 & 0.76 & 0.10 & 0.17 & 0.76 & 0.44 & 0.42 \\ \cline{2-29}
        & \cellcolor{verylightgray} Avg.
        & \cellcolor{verylightgray} \textbf{0.54} 
        & \cellcolor{verylightgray} \textbf{0.22}
        & \cellcolor{verylightgray} \textbf{0.26}
        & \cellcolor{verylightgray} \textbf{0.44}
        & \cellcolor{verylightgray} \textbf{0.24}
        & \cellcolor{verylightgray} \textbf{0.28}
        & \cellcolor{verylightgray} \textbf{0.59}
        & \cellcolor{verylightgray} \textbf{0.42}
        & \cellcolor{verylightgray} \textbf{0.46}
        & \cellcolor{verylightgray} \textbf{0.58}
        & \cellcolor{verylightgray} \textbf{0.14} 
        & \cellcolor{verylightgray} \textbf{0.19}
        & \cellcolor{verylightgray} \textbf{0.59}
        & \cellcolor{verylightgray} \textbf{0.19}
        & \cellcolor{verylightgray} \textbf{0.23}
        & \cellcolor{verylightgray} \textbf{0.61}
        & \cellcolor{verylightgray} \textbf{0.34} 
        & \cellcolor{verylightgray} \textbf{0.36}
        & \cellcolor{verylightgray} \textbf{1.00}
        & \cellcolor{verylightgray} \textbf{0.04}
        & \cellcolor{verylightgray} \textbf{0.08}
        & \cellcolor{verylightgray} \textbf{0.80}
        & \cellcolor{verylightgray} \textbf{0.12}
        & \cellcolor{verylightgray} \textbf{0.20} 
        & \cellcolor{verylightgray} \textbf{0.76}
        & \cellcolor{verylightgray} \textbf{0.33}
        & \cellcolor{verylightgray} \textbf{0.38}
        \\ \hline
        \multirow{6}{*}{\rotatebox{90}{Dataset $D_2$}} & R1 & 0.26 & 0.14 & 0.14 & 0.24 & 0.19 & 0.17 & 0.27 & 0.21 & 0.23 & 0.33 & 0.10 & 0.13 & 0.27 & 0.08 & 0.11 & 0.33 & 0.17 & 0.19 & 0.54 & 0.04 & 0.07 & 0.38 & 0.17 & 0.20 & 0.28 & 0.17 & 0.17 \\ \cline{2-29}
        & R2 & 0.33 & 0.13 & 0.17 & 0.36 & 0.25 & 0.25 & 0.25 & 0.30 & 0.23 & 0.24 & 0.13 & 0.11 & 0.33 & 0.10 & 0.13 & 0.23 & 0.14 & 0.16 & 0.44 & 0.06 & 0.09 & 0.44 & 0.16 & 0.20 & 0.26 & 0.14 & 0.15 \\ \cline{2-29}
        & R3 & 0.20 & 0.09 & 0.10 & 0.25 & 0.11 & 0.14 & 0.24 & 0.24 & 0.21 & 0.20 & 0.07 & 0.09 & 0.35 & 0.19 & 0.18 & 0.26 & 0.19 & 0.16 & 0.50 & 0.07 & 0.10 & 0.41 & 0.14 & 0.18 & 0.25 & 0.14 & 0.14 \\ \cline{2-29}
        & R4 & 0.13 & 0.04 & 0.06 & 0.31 & 0.21 & 0.21 & 0.25 & 0.29 & 0.23 & 0.27 & 0.07 & 0.09 & 0.21 & 0.05 & 0.07 & 0.24 & 0.17 & 0.14 & 0.52 & 0.04 & 0.07 & 0.44 & 0.16 & 0.20 & 0.24 & 0.13 & 0.12 \\ \cline{2-29}
        & R5 & 0.18 & 0.07 & 0.09 & 0.28 & 0.15 & 0.18 & 0.25 & 0.30 & 0.24 & 0.22 & 0.06 & 0.07 & 0.19 & 0.04 & 0.06 & 0.23 & 0.12 & 0.15 & 0.60 & 0.06 & 0.10 & 0.43 & 0.17 & 0.22 & 0.16 & 0.06 & 0.08 \\ \cline{2-29}
  &\cellcolor{verylightgray} Avg. & \cellcolor{verylightgray}\textbf{0.22} & \cellcolor{verylightgray}\textbf{0.09} & \cellcolor{verylightgray}\textbf{0.11} & \cellcolor{verylightgray}\textbf{0.29} & \cellcolor{verylightgray}\textbf{0.18} & \cellcolor{verylightgray}\textbf{0.20} & \cellcolor{verylightgray}\textbf{0.26} & \cellcolor{verylightgray}\textbf{0.27} & \cellcolor{verylightgray}\textbf{0.23} & \cellcolor{verylightgray}\textbf{0.26} & \cellcolor{verylightgray}\textbf{0.08} & \cellcolor{verylightgray}\textbf{0.10} & \cellcolor{verylightgray}\textbf{0.27} & \cellcolor{verylightgray}\textbf{0.09} & \cellcolor{verylightgray}\textbf{0.11} & \cellcolor{verylightgray}\textbf{0.26} & \cellcolor{verylightgray}\textbf{0.16} & \cellcolor{verylightgray}\textbf{0.16} & \cellcolor{verylightgray}\textbf{0.54} & \cellcolor{verylightgray}\textbf{0.05} & \cellcolor{verylightgray}\textbf{0.08} & \cellcolor{verylightgray}\textbf{0.42} & \cellcolor{verylightgray}\textbf{0.16} & \cellcolor{verylightgray}\textbf{0.19} & \cellcolor{verylightgray}\textbf{0.24} & \cellcolor{verylightgray}\textbf{0.13} & \cellcolor{verylightgray}\textbf{0.14} \\ \midrule

        \multirow{6}{*}{\rotatebox{90}{Dataset $D_{m05}$}} & R1 & 0.52 & 0.21 & 0.25 & 0.36 & 0.17 & 0.21 & 0.34 & 0.29 & 0.29 & 0.56 & 0.13 & 0.19 & 0.48 & 0.14 & 0.20 & 0.37 & 0.30 & 0.27 & 0.54 & 0.04 & 0.07 & 0.63 & 0.07 & 0.12 & 0.33 & 0.11 & 0.14 \\ \cline{2-29}
        & R2 & 0.33 & 0.10 & 0.15 & 0.23 & 0.09 & 0.12 & 0.24 & 0.26 & 0.20 & 0.47 & 0.12 & 0.14 & 0.30 & 0.06 & 0.09 & 0.32 & 0.26 & 0.22 & 0.44 & 0.06 & 0.09 & 0.56 & 0.07 & 0.11 & 0.46 & 0.18 & 0.25 \\ \cline{2-29}
        & R3 & 0.39 & 0.13 & 0.18 & 0.51 & 0.16 & 0.22 & 0.28 & 0.22 & 0.24 & 0.41 & 0.08 & 0.11 & 0.26 & 0.15 & 0.11 & 0.35 & 0.18 & 0.17 & 0.50 & 0.07 & 0.10 & 0.50 & 0.06 & 0.10 & 0.48 & 0.13 & 0.20 \\ \cline{2-29}
        & R4 & 0.36 & 0.20 & 0.22 & 0.34 & 0.19 & 0.23 & 0.27 & 0.25 & 0.23 & 0.33 & 0.03 & 0.05 & 0.32 & 0.06 & 0.09 & 0.38 & 0.20 & 0.24 & 0.52 & 0.04 & 0.07 & 0.63 & 0.10 & 0.15 & 0.39 & 0.14 & 0.14 \\ \cline{2-29}
        & R5 & 0.36 & 0.10 & 0.15 & 0.33 & 0.15 & 0.18 & 0.32 & 0.32 & 0.28 & 0.33 & 0.02 & 0.04 & 0.36 & 0.12 & 0.13 & 0.38 & 0.19 & 0.21 & 0.60 & 0.06 & 0.10 & 0.52 & 0.11 & 0.17 & 0.46 & 0.26 & 0.28 \\ \cline{2-29}
       & \cellcolor{verylightgray} Avg. & \cellcolor{verylightgray}\textbf{0.39} & \cellcolor{verylightgray}\textbf{0.13} & \cellcolor{verylightgray}\textbf{0.18} & \cellcolor{verylightgray}\textbf{0.36} & \cellcolor{verylightgray}\textbf{0.15} & \cellcolor{verylightgray}\textbf{0.18} & \cellcolor{verylightgray}\textbf{0.29} & \cellcolor{verylightgray}\textbf{0.25} & \cellcolor{verylightgray}\textbf{0.25} & \cellcolor{verylightgray}\textbf{0.43} & \cellcolor{verylightgray}\textbf{0.07} & \cellcolor{verylightgray}\textbf{0.10} & \cellcolor{verylightgray}\textbf{0.32} & \cellcolor{verylightgray}\textbf{0.11} & \cellcolor{verylightgray}\textbf{0.13} & \cellcolor{verylightgray}\textbf{0.36} & \cellcolor{verylightgray}\textbf{0.23} & \cellcolor{verylightgray}\textbf{0.20} & \cellcolor{verylightgray}\textbf{0.54} & \cellcolor{verylightgray}\textbf{0.05} & \cellcolor{verylightgray}\textbf{0.08} & \cellcolor{verylightgray}\textbf{0.59} & \cellcolor{verylightgray}\textbf{0.08} & \cellcolor{verylightgray}\textbf{0.13} & \cellcolor{verylightgray}\textbf{0.38} & \cellcolor{verylightgray}\textbf{0.18} & \cellcolor{verylightgray}\textbf{0.21} \\ \hline

    \end{tabular}}     
       \label{tab:attr_results}
\end{table*}

A similar trend in performance accuracy can be observed for recommending \textit{attribute} operations summarized in Table \ref{tab:attr_results}. As expected, the accuracy is lower compared to class operations, \ie the best F1-score value is 0.46 using configuration $C_{1.3}$. This result can be explained by the higher variability in defining modeling operations. Concerning the impact of synthetic operations, the results confirm that $D_1$ offers better performance, even though the delta between the synthetic data is lower than the class recommendations. It is worth mentioning that the obtained performance is in line with state-of-the-art IMAs \cite{di2022memorec,weyssow2022recommending,chaaben2023towards} used in modeling completion tasks. Furthermore, we report that the NEMO approach \cite{di2022finding}, the most relevant approach to ours, achieves 0.60 of accuracy on a curated dataset. In this respect, our approach employs traces extracted from real-world models. 

Similarly to $RQ_1$, we evaluated the time needed to compute the training and testing phase considering only the \MG tool, thus excluding the time to generate the traces. Overall, the time to load and encode the training traces is equal to 0.07 seconds on average for each fold, while 7 seconds are required to perform the recommendation phase.

\textbf{Answer to RQ$_2$:} The results show that real traces are better to feed traditional IMAs like \MG. However, synthetic traces can be used to augment the training set in a faster way and preserve the accuracy of other state-of-the-art tools.

\subsubsection*{\textbf{Addressing $RQ_3$}} To answer this question, we run \MG on Dataset $D_{3}$ using the previous datasets as training, \ie $D_1$, $D_2$, and $D{m05}$. In addition, we derive two additional datasets by mixing different ratios of synthetic and real traces, \ie $D{m02}$ and $D{m08}$, where the ratio of synthetic traces are 0.2 and 0.8 out of the total number, respectively. 
\begin{figure}[ht!]
    \centering
     \subfloat[Attributes results on $D_{3}$.]{%
    \includegraphics[width=0.45\linewidth]{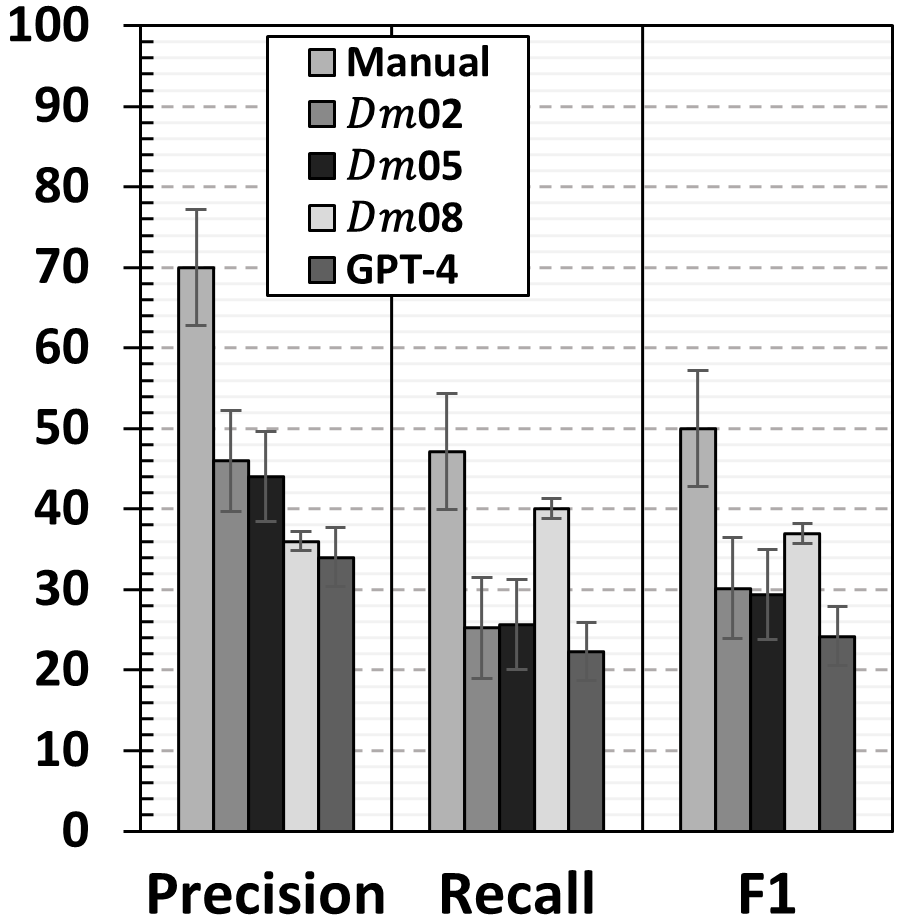}}
    \label{fig:attr_res_d3} \
      \subfloat[Class results on $D_{3}$.]{%
       \includegraphics[width=0.45\linewidth]{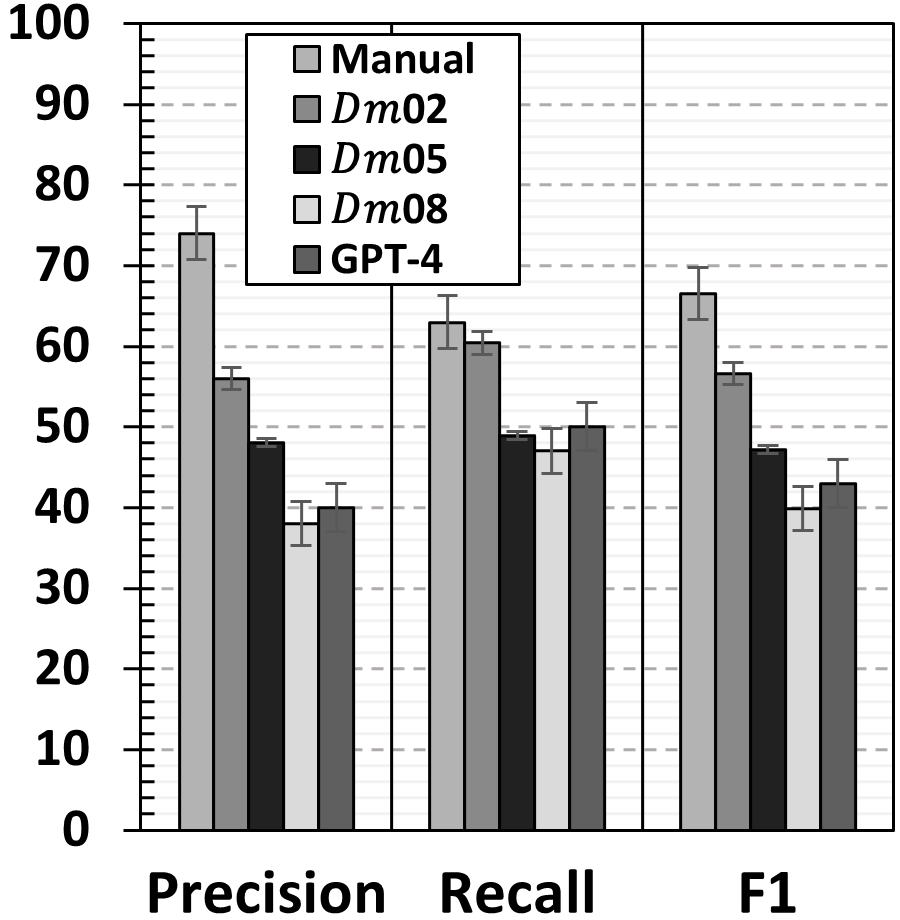}}
    \label{fig:class_res_d3} 
  \caption{Validation results.}
  \label{fig:res_d3_final} 
\end{figure}
Concretely, we used $D_3$ datasets as the \textit{validation set} of our approach, aiming to evaluate how the produced data can be used in different application domains. Figure \ref{fig:res_d3_final} (a) and Figure \ref{fig:res_d3_final} (b) depict the results obtained using the three training sets for class and attribute recommendations, respectively. Notably, we used configuration C3.3 to compute the results, which leads to better performance, as shown in the previous research question. While the human-based traces are still the best training set for \MG, the usage of synthetic data can be used to recommend operations in different industrial contexts, \ie the metrics values are in line with the ones obtained in the previous RQ. In addition, we see that the operations generated by the GPT-4 model tend to increase the recall values while human-generated traces lower the precision values on average. Therefore, mixing human and synthetic traces can represent an adequate trade-off to reduce the number of false negatives. This is confirmed by analyzing the results obtained with the novel datasets, \ie $D{m02}$ and $D{m08}$. Concerning the time needed for the recommendations, it is worth noticing that the training time is reduced from 7 to 2 seconds on average as we are using \MG pre-trained weights to perform the recommendations activity. 
%
%
In addition, we demonstrate that our approach can be used in different application contexts, thus representing a suitable alternative to cope with \textbf{CH3} discussed in Section \ref{sec:motivationalexample}. Concretely, IMAs can be trained with synthetic traces that are not identical but similar to the target ones, thus overcoming privacy issues in an industrial context.

\textbf{Answer to RQ$_3$:} Even though we report a degradation of performances when using the pre-trained \MG model, our approach can be used to support the specification of IMAs when training data are missing. In practice, the recommendations that are produced might be useful in different application domains.
%



\section{Threats To Validity}
\label{sec:Threats}
This section discusses the threats that may hamper the results of our work and the mitigations. 
Threats \textit{internal validity} concerns two main aspects, \ie the evaluation of synthetic data generated using LLMs and the experiments conducted to evaluate the IMA component. Concerning the generated data, hallucination may lead to incorrect operations, thus feeding the IMA component with unsuitable data. We mitigate this threat by experimenting with four different popular LLMs and evaluating the generated data using well-founded metrics. In addition, we adapt the concept of hallucination to the modeling context, focusing on generating only additive events, \ie we did not consider removing operations to reduce any bias in computing the proposed hallucination metric. Regarding the IMA accuracy, the computed metrics can lead to incorrect results. To cope with this issue, we design three different configurations to simulate different levels of completion, resembling the evaluation setting exploited by state-of-the-art tools. Furthermore, we confirmed the outcomes of the 5-fold cross-validation using well-known statistical indexes.

Concerning the \textit{external validity}, the main issue is the generalizability of the proposed framework, \ie the obtained results may vary considering a different set of modeling tools. To mitigate this, we validate the proposed approach by using a well-founded modeling component for each conceptual block of the architecture, \ie the modeling environment, the trace recorder, and the IMA assistant. Furthermore, we employ an additional modeling dataset exploited in several EU projects. In such a way, we validate synthetic traces in different domain applications as discussed in $RQ_3$. 


\section{Conclusion}
\label{sec:Conclusion}
This paper proposes a conceptual framework to support automated activities in the context of MBSE leveraging modeling event recorders, intelligent modeling assistants, and large language models. In particular, we used prominent LLMs to generate synthetic traces using an in-context few-shots prompt engineering strategy, aiming at resembling human-style operations. The findings of the study demonstrate that LLMs can be used to generate traces in a specific format even though the evaluated assistant suffers from degradation of performance when delivering recommendations. Nonetheless, generating modeling operations can be seen as a valuable alternative when training data are not available due to different factors, \eg internal regulations or privacy issues.

In future works, we plan to extend the evaluation to additional application contexts. In addition, we will include
different modeling tools to validate the quality of synthetic traces. Last but not least, we plan to fully automate the whole pipeline and collect quality feedback from modelers.


\section*{Acknowledgments}
This work was supported by the AIDOaRt project grant from the ECSEL Joint Undertaking (JU) (grant n. 101007350). Furthermore, we thank Volvo Construction Equipment and TEKNE companies. 








\bibliographystyle{acm}
\bibliography{references}

\end{document}